\documentclass[draft,onecolumn,12pt]{IEEEtran} % Pablo

\usepackage[cmex10]{amsmath}
\usepackage{cases}
\usepackage{amsfonts}	%permite hacer R y N como corresponde
\usepackage{cite}
\usepackage{color}
\usepackage{graphicx}	
\usepackage{amssymb,url}
\usepackage{array}	
\usepackage{dsfont}
\usepackage{pgfplots,stfloats}
\usepackage[T1]{fontenc}
\graphicspath{{./Figs/}}                   
\usepackage[normalem]{ulem}
\usetikzlibrary{positioning,backgrounds}
\newcounter{MYtempeqncnt}

\newtheorem{lemma}{Lemma}[section]
\newtheorem{theorem}{Theorem}[section]

\newtheorem{remark}{Remark}[section]
\newtheorem{definition}{Definition}[section]
\newcommand{\Ex}{\mathbb{E}}
\newcommand{\prob}{\mathbb{P}}
\newcommand{\Aset}{\mathcal{A}}
\newcommand{\Sset}{\mathcal{S}}
\newcommand{\Ti}{\mathcal{T}_i}
\newcommand{\Mi}{\mathcal{M}_i}
\newcommand{\avec}{\mathbf{A}}
\newcommand{\mvec}{\mathbf{m}}
\newcommand{\vvec}{\mathbf{V}}
\newcommand{\rvec}{\mathbf{D}}
\newcommand{\uvec}{\mathbf{S}}
\newcommand{\hvec}{\mathbf{H}}

\newcommand{\R}{\mathbb{R}}
\newcommand{\bigone}{\mathds{1}}
\newcommand{\B}{\mathcal{B}}
\newcommand{\M}{\mathcal{M}}
\newcommand{\K}{\mathcal{K}}
\newcommand{\T}{\mathcal{T}}
\newcommand{\C}{\mathcal{C}}
\newcommand{\Oi}{\mathcal{O}_i}

\pgfplotsset{tick label style={font=\scriptsize},
			legend style = {font=\small},
			xlabel style ={font = \small},
			ylabel style ={font = \small},
			xticklabel style={
			/pgf/number format/precision=1	,
			/pgf/number format/fixed},
			yticklabel style={
			/pgf/number format/precision=2,
			/pgf/number format/fixed,
			/pgf/number format/fixed zerofill},
			legend style = {cells={anchor=west}},
			grid = both,
			every axis plot/.append style={line width=1pt}
			}

\begin{document}
\title{On Fundamental Trade-offs of Device-to-Device Communications in Large Wireless Networks}

\author{Andr\'es~Altieri,
        Pablo~Piantanida,
        Leonardo~Rey~Vega,
        and Cecilia~G.~Galarza 
\thanks{Partially supported by the Eiffel scholarship of France, project UBACyT 2002013100751BA, and FP7 Network of Excellence in Wireless Communications NEWCOM\#. 

P. Piantanida and Andrés Altieri are with the Laboratoire des Signaux et Systèmes (L2S, UMR8506) CentraleSupélec-CNRS-Université Paris-Sud, F-91190 Gif-sur-Yvette, France (e-mail: \{andres.altieri, pablo.piantanida\}@centralesupelec.fr).

L. Rey Vega and C. G. Galarza are with the Departments of Electronics (Universidad de Buenos Aires) and CSC-CONICET, Buenos Aires, Argentina (e-mail: \{lrey, cgalar\}@fi.uba.ar).}
}

%Thanks to ADD -->   The work of P. Piantanida was partially supported by  the Celtic European project SHARING.  
%\author{\IEEEauthorblockN{Andr\'{e}s Altieri}
%\IEEEauthorblockA{Dept. of Telecommunications\\
%SUPELEC\\
%Gif-sur-Yvette, France\\
%Email: aaltieri@fi.uba.ar}
%\and
%\IEEEauthorblockN{Pablo Piantanida}
%\IEEEauthorblockA{Dept. of Telecommunications\\
%SUPELEC\\
%Gif-sur-Yvette, France\\
%Email: pablo.piantanida@supelec.fr}
%\and
%\IEEEauthorblockA{Leonardo Rey Vega, Cecilia G. Galarza}
%\IEEEauthorblockA{School of Engineering \\
%University of Buenos Aires and CONICET\\
%Buenos Aires, Argentina\\
%Email: \{lrey, cgalar\}@fi.uba.ar}

\maketitle

\begin{abstract}
This paper studies the gains, in terms of served requests, attainable through out-of-band device-to-device (D2D) video exchanges in large cellular networks. A stochastic framework, in which users are clustered to exchange videos, is introduced, considering several aspects of this problem: the video-caching policy, user matching for exchanges, aspects regarding scheduling and transmissions. A family of \emph{admissible protocols} is introduced: in each protocol the users are clustered by means of a hard-core point process and, within the  clusters, video exchanges take place.
Two metrics, quantifying the ``local'' and ``global'' fraction of video requests served through D2D are defined, and relevant trade-off regions involving these metrics, as well as quality-of-service constraints, are identified. A simple communication strategy is proposed and analyzed, to obtain inner bounds to the trade-off regions, and draw conclusions on the performance attainable through D2D. To this end, an analysis of the time-varying interference that the nodes experience,  and tight approximations of its Laplace transform are derived. 
\end{abstract}

\section{Introduction}
\label{sec:intro}

\subsection{Motivation and Related Work}
Cellular device-to-device (D2D) communications, in which two or more mobile users establish a direct link without going through the Base Station (BS), have emerged as a viable alternative to partially cope with the increasing  requirements %demands in data traffic and connectivity 
that cellular networks will face in the future.  Generally speaking, D2D communications are opportunistic, one-hop, short range transmissions in which the BS can be used for coordination and acts as a last case fall-back alternative \cite{LinAndGhoRat13}. This allows, for a higher spatial frequency reuse, energy efficiency, coverage extension, and a reduced backhaul load. 
%The presence of the BS is an important advantage over other network topologies such as \emph{ad hoc} networks, which rely solely on themselves for coordination, transmission and routing \cite{LinAndGhoRat13}.   
The scope of D2D communications is very wide, from machine-to-machine, gaming and relaying, to content distribution, and public safety networks~\cite{LinAndGhoRat13}.
Among these, an important application is video content distribution. 
This is because, in the next few years, traffic from wireless and mobile devices will exceed traffic from wired devices, and this will be largely related to an increase in \emph{video on demand} (VoD) and Internet-to-TV downloads~\cite{Cisco2013}.
The asynchronous nature of VoD requests implies that, in many cases, multicasting strategies for video transmissions cannot be employed, even though a small library of videos may be accessed by many users~\cite{CaireMolischJi2013}. A recent approach to mitigate this consists in including small distributed caching stations with a limited backhaul that can locally serve video demands \cite{GMDC_2013,SGDMC_DistCach2013}. Another approach~\cite{CaireMolischJi2013} is to take advantage of the unused storage space available in many wireless devices to store and exchange videos locally. For example, a user may keep its watched videos to satisfy nearby requests, or certain videos could be cached during moments of low network load.
In this paper we focus on the second approach, which does not require dedicated storage units. Our main goal is to study the potential benefits achievable through a distributed user-caching strategy, by considering the fraction of mean video requests that could be served through D2D, without requiring the BS to transmit them. This may yield some insight on the impact of D2D in terms of video availability and backhaul load, which may have implications both economically and in terms of quality of service. To this end, we introduce a simple framework for analysis, based on a stochastic geometry model \cite{stochastic_geometry2009,BB2010}. In this framework, users are assumed to be grouped into clusters where D2D video exchanges take place. We attempt to consider the problem of establishing matches between requesting and caching users, and the problem of scheduling and transmitting, involving slow fading, path loss, and interference between nodes. We focus on \emph{out-of-band} D2D, which uses 
bands outside the cellular ones, increasing the frequency reuse and mitigating interference in the cellular band. We also study the trade-offs between the fraction of requests  served ``locally'' (per cluster) and ``globally'' (in an arbitrary region) in the  network. 
%the ``global fraction'' is an overall indication of how beneficial D2D can be, but this does not necessarily mean that, in specific areas, performance levels are as good as expected. 

Some related works which focus on D2D through local distributed caching include~\cite{CaireMolischJi2013,SGDMC_DistCach2013,GolrezaeiDimakisMolisch2012,JCM_2013,CaireMolischJi2013_D2DC,CaireMolischJi2014}. These works consider finite area networks 
with a fixed number of users, distributed uniformly or on a regular grid.
%in which a fixed number of users are distributed either uniformly or on a regular grid. 
The model for transmission failure is generally the distance-based \emph{protocol model} introduced in \cite{gupta_capacity_2000}. They also consider \emph{out-of-band} D2D but focus on the optimal asymptotic scaling laws of the networks. For example, in \cite{CaireMolischJi2013,JCM_2013} a one-hop network in which users are clustered and cache videos is studied with this model, and a throughput-outage trade-off is characterized for various regimes, in terms of scaling laws as the number of nodes and the library size grow to infinity. In \cite{GolrezaeiDimakisMolisch2012} they find an optimal collaboration distance to balance interference, and analyze the scaling behavior of the benefits of D2D. On the other hand, our approach considers an  infinite-area constant-density model in which transmissions are impaired by path loss and fading. 

%On the other hand, in  \cite{SGDMC_DistCach2013} they consider 
%We analyze the optimum way of assigning
%files to the helpers, in order to minimize the expected
%downloading time for files. We distinguish between the
%uncoded case (where only complete files are stored) and
%the coded case, where segments of Fountain-encoded
%versions of the video files are stored at helpers.

Other works on D2D through stochastic geometry models are \cite{YeAndd2dhop13,YeAndd2d13,LinAndGhoJSAC13,Bastug2015}. In general, these works are not focused on video distribution, which we attempt to analyze, but on general traffic and general aspects of D2D communications. Hence, they do not consider the problems of user matching, user request statistics and caching policies, which become central in the video distribution problem. In \cite{YeAndd2dhop13}, the authors study the optimal downlink spectrum partition between D2D and BS transmissions. In \cite{YeAndd2d13}, the authors analyze a D2D\emph{ in-band overlaid} cellular network model and find expressions for several performance metrics. Finally, \cite{Bastug2015} considers the problem of video distribution through distributed storage BSs.

\subsection{Main Contributions}
The main goal of this paper is to study the number of requests that could  be served by D2D instead of asking the BS for  a transmission. 
To this end, we propose a stochastic geometry framework, with the following characteristics:
\begin{itemize}
\item Requesting users (destinations) and cooperative users (with cached videos) are  distributed in space as a Poisson  point process (PP). Transmissions are affected by path loss, slow fading and interference.
\item Users are grouped in disjoint clusters where D2D exchanges take place. A family of \emph{admissible protocols} is introduced: each  protocol is composed of a \emph{clustering strategy} induced by any hard-core PP \cite{stochastic_geometry2009}, and any suitable \emph{in-cluster communication  strategy}, which dictates the communication schemes of users inside the clusters.
\end{itemize}
In this setting, we define two metrics of interest, which characterize the performance of D2D in terms of served requests:
\begin{itemize}
\item  A \emph{global} metric, that measures the ratio between the spatial density of served requests and the total density of requests. This gives the global fraction of the video requests which could be served through D2D exchanges without using the downlink of the cellular network.
\item A \emph{local} metric, that measures the ratio between the average number of served requests and total requests in a cluster. This can be interpreted as an indication of what gains could be expected in a localized region in space, in which certain level of service is required.
\end{itemize}
Although these metrics address the three aspects of the problem mentioned earlier, introducing a link-quality constraint is reasonable to model the delay constraints which may be required in video distribution. %Otherwise, a large fraction of requests could be satisfied through low-rate transmissions, i.e. having a large delay, which may be not be reasonable in this setup.
% but a link-quality requirement has to be considered in order to make the transmission problem involved meaningful. %These metrics become relevant when quality-of-service or network-design requirements are considered. 
For this reason, we introduce three trade-off regions pertaining these metrics:
\begin{itemize}
\item The \emph{global metric-average rate} trade-off region, which pertains the fraction of requests than can be served considering an average rate
%\footnote{We assume the transmission block is long enough to consider the notion of reliable communication, but shorter than the dynamics of the slow fading process, whose realizations may vary between video streams transmitted over different blocks. An average delay constraint can guarantee an application requirement, e.g., video transmission might be less delay tolerant than voice.
requirement over the cluster.
\item The \emph{global metric-average rate and cluster density} trade-offs region, which refers to the local fraction of requests that can be served considering that an average rate and a minimal cluster density are required.
\item The \emph{global-local} trade-off regions, which pertain the balance between the global and the local density of served requests which are attainable simultaneously.
\end{itemize}
Determining these regions implies characterizing  the optimal communication scheme among the family of admissible protocols mentioned before. Since optimal communication schemes remain unknown  for each trade-off region, we analyze a simple in-cluster communication  strategy, which can be paired with any clustering strategy to obtain a protocol. This will give an inner bound to the trade-off regions. In this strategy, users  which request videos and those with cached videos are paired and a one-hop transmission takes place. Interference within clusters is avoided by precluding simultaneous transmissions through a  time-division multiple access (TDMA) scheme which shares the time resource between transmitters. It is shown that the TDMA scheme implies that a user will experience a time-varying interference during a slot. An analysis which takes this into account is performed in order to determine the rates that a user can achieve. This analysis, which is not usually considered in a stochastic geometry setup, may be of interest for scenarios other than D2D. We then evaluate the global and local metrics for all the protocols obtained by pairing this in-cluster communication  strategy  with any clustering strategy. In this way, we obtain a set of inner bounds to the optimal trade-off regions, which give an indication of the possible gains  through D2D.
%We then evaluate the global and local metrics for this protocol, which provide inner bounds to the optimal trade-off regions, and give an indication of the possible gains  through D2D. 
Finally, we numerically evaluate these inner bounds in different scenarios, considering the clustering strategy induced by a type II Matérn hard-core PP and the translated grid PP~\cite{stochastic_geometry2009}, under a Rayleigh fading model and a lognormal shadow fading model in which line of sight (LOS) may be present inside the clusters.  In the Matérn hard-core PP with Rayleigh fading, we also develop approximations to the Laplace transform (LT)  of the interference anywhere in a cluster. Through this analysis, we draw conclusions regarding the performance of D2D communications in cellular networks.
%Furthermore, these approximations on the LT could also be used in other scenarios in which type II hard-core processes are considered, e.g., in carrier-sense-multiple-access (CSMA) or cell-based networks in which the impact of user placement within the cell is to be studied. 

The paper is organized as follows. In Section \ref{sec:model}, we introduce the network model, the family of admissible protocols and the main metrics and trade-offs under study. In Section \ref{sec:stratan}, we introduce and analyze a simple communication strategy. % and derive the approximations on the LT of the interference.
In Section \ref{sec:plots}, we present some plots and comments. In Section \ref{sec:discus} we discuss our findings, and proofs are  in the Appendix.

\subsubsection*{Notation} 
$F_X(\cdot)$ is the cumulative distribution function (cdf) of the random variable (RV) $X$ and $\bar{F}_X(\cdot)$ is its complementary cdf.  $\B(x,y)$ is the ball of radius $y$ centered at $x$. $\bigone_A$ is the indicator for an event $A$. All logarithms are to base two unless specified, $\C(x) \triangleq \log(1+x)$, $(x)^+ \triangleq \max\{x,0\}$.

\section{System model and admissible protocols} \label{sec:model}
%In Section \ref{sec:userdesc} we introduce a simple model of how users are distributed and cache videos. Also, we introduce a family of clustering strategies, which group the users into clusters, in which D2D exchanges take place. %This spatial model is fully characterized through an independently marked point process. 
%In Section \ref{sec:inclust} we define a family of in-cluster coordination strategies, which define how users may interact inside the cluster to exchange videos. In Section \ref{sec:metrics} we introduce a set of metrics which allow us to study the performance attainable through the set of clustered D2D strategies. %We refer to  videos as videos, although  we focus  on the former. 
\subsection{Clustering Strategies and Spatial Model} \label{sec:userdesc}
%We now introduce a basic spatial model in which users may attempt to exchange videos through D2D. 
%Our goal is to introduce a basic scenario in which D2D file exchanges can take place between users which have files cached and others who request them.
%We refer to  videos generically as files, although our focus is set on the former. 
%Throughout the paper we refer to the videos generically as files, although we focus on the former. 
We consider an infinite planar network in which:
\begin{itemize}
\item Users who request videos are distributed according to an homogeneous Poisson PP $\Phi_r$, of intensity $\lambda_r$.  Users who cache videos are distributed according to an homogeneous Poisson PP $\Phi_u$, of intensity $\lambda_u$, independent of $\Phi_r$\footnote{We can consider these two PPs as originating from a single Poisson PP of intensity $\lambda_u+\lambda_r$, where then a user decides to become a requesting user with probability $\lambda_r/(\lambda_u+\lambda_r)$, independent of everything else, and the rest are caching users. This separation is done to simplify the model.}. %In reality, requesting users would also cache videos,  increasing the likelihood of finding videos and the number of served requests (provided an efficient medium access and transmission strategy is considered).}.
Users attempt to exchange videos through D2D to reduce the load on the downlink of the cellular network, and do so outside the downlink  band  so there is no interference between cellular and D2D communications.
\item Each user in $\Phi_r$ requests a video, which is selected independently according to a discrete distribution $p_V(v)$, where $1\leq v \leq L$, and $L$ is the library size. 
In numerical results, we assume that the videos are sorted according to their popularity, which implies that $p_V(v)$ is the probability of requesting the $v$-th most popular video. This distribution is commonly~\cite{CaireMolischJi2013,SGDMC_DistCach2013,GolrezaeiDimakisMolisch2012,JCM_2013} assumed to be a Zipf distribution of parameter $0 < \gamma < 1$, 
% This distribution is commonly taken to be a Zipf distribution 
%In numerical results, we assume files are requested according to their popularity distribution $p_V$ which  follows 
%In numerical results we assume $p_V$ to be a Zipf distribution with parameter $0 < \gamma < 1$, that is:
%\begin{equation}
$p_V(v) = \frac{v^{-\gamma}}{\sum_{i = 1}^L i^{-\gamma}}.$ %\ \ \ v\in\{1, \dots, L\}.
%\end{equation}  
\item Each user of $\Phi_u$ has $M$ (fixed) videos cached, which, for simplicity, are selected independently according to a discrete distribution $p_A(a)$, $1 \leq a \leq L$.
%\sout{, where the support of the distribution $L_A \leq L$ denotes the number of videos that may be cached out of the total number of videos in the library. A value $L_A < L$ could be selected to allow the users to cache only a certain subset of videos, for example, the $L_A$ most popular ones.  }
For numerical results we assume that $p_A\equiv p_V$; this can be motivated assuming that users cache videos they  watch.
%\sout{, with $L_A=L$.}
%\item \sout{Transmissions are subject to both slow fading and path loss. The power received at $y$ by a unit-power transmission from $x$ is $|h_{xy}|^2 l(x,y)$ where $l(x,y)= \|x-y\|^{-\alpha}$ ($\alpha >2$) is the usual path loss function and $|h_{xy}|^2$ is the power gain of fading with unit mean. }
%\sout{In our analysis we focus on the interference generated between the nodes using D2D, and focus on the signal-to-interference ratio (SIR). Independent heterogeneous background interference or noise could be added in a straightforward manner.}
\end{itemize}
To exchange videos, users of $\Phi_r$ and $\Phi_u$ are grouped into disjoint clusters which, for simplicity, are approximated as discs of radius $R_c$.
The users of $\Phi_r$ which are not clustered will ask for the videos directly to the BS, while the users who are clustered can search in their cluster for a user from $\Phi_u$ who has the video, and request a transmission through D2D. 
%A request by a user is said to be \emph{served} if some other user has the requested video and the transmission is successful. In what follows we describe this family of protocols.
Assuming that the clusters are disjoint implies there is a minimum distance of at least $2R_c$ between their centers, so we can model the spatial distribution of cluster centers as a
hard-core PP \cite{stochastic_geometry2009}, which guarantees this clearance. 
%Assuming that the disk-shaped clusters do not intersect implies there is a minimum distance of at least $2R_c$ between the  centers and hence, modeling the cluster centers as a hard-core PP \cite{stochastic_geometry2009}, which guarantees this clearance, is reasonable. 
This leads to the following definition.
%With each  hard-core process we may define a process of clustered users:
\begin{definition}[Process of clustered users] \label{def:parentp}
Given a cluster radius $R_c > 0$, a PP of clustered users $\Phi_c$ is constructed from $\Phi_u$ and $\Phi_r$  as follows:
\begin{equation}
\Phi_c = \bigcup_{x \in \Phi_p} \B(x,R_c) \cap (\Phi_r \cup \Phi_u), \label{eq:phic1}
\end{equation}
where $\Phi_p = \{x_i\}$ is a stationary \emph{parent} hard-core PP  of intensity $\lambda_p > 0$ and clearance  $\delta \geq 2R_c$.
\end{definition}
Any hard-core stationary PP with clearance $\delta \geq 2R_c$ will generate a cluster PP. In Fig. \ref{fig:netrep} we can see a representation of the network. 
\begin{remark} \label{req:equivmod}
The same model is obtained if we first deploy the discs of radius $R_c$ with centers in $\Phi_p$, then create independent Poisson PPs of the same intensities as $\Phi_u$ and $\Phi_r$ inside and outside the discs.
\end{remark}

We consider an attenuation model with slow fading and path loss, 
with possibly a different attenuation model for transmissions inside a cluster and one between clusters.
This is for scenarios in which transmitters inside a cluster are collocated or that LOS is present. For this we consider:
\begin{itemize}
\item \emph{Inter-cluster attenuation model:} a transmission of power $P$ from $x$ to $y$  in different clusters is received with power:
\begin{equation}
P |h_{x,y}|^2 l(x,y),  \label{eq:interc}
\end{equation} with $|h_{xy}|^2$  a  fading coefficient, independent of everything, and $l(x,y) \equiv l(||x-y||)$ is a path loss function.
\item \emph{Intra-cluster attenuation model: }a transmission of power $P$ from $x$ to $y$ in the same cluster is received with power $P |g_{xy}|^2$, where $|g_{xy}|^2$ is a random power attenuation coefficient which contains fading and path loss, and whose distribution depends on $||x-y||$. 
\end{itemize}
We focus on the interference generated between the nodes using D2D, and on the signal-to-interference ratio (SIR). Independent  background interference or noise could be added
directly. 
Each cluster will have a family of associated users which cache or request videos, which will be the points of the original PPs $\Phi_u$ and $\Phi_r$ which fall in the discs. Each of these users will be represented by some information, mainly, their positions, the video(s) they cache or request, and fading coefficients towards other users. 
This information is represented as a vector of RVs, associated to each cluster center, and, according to the assumptions stated earlier, independent among clusters. So we can  represent the network of clusters as a stationary independently marked PP \cite{stochastic_geometry2009,BB2010} 
%We can represent  this information  as a vector of RVs which is associated to the point of each cluster center, and, due to the assumptions we have stated  earlier in the section, these marks will be independent among clusters. This means we can represent the network of clusters as a stationary independently marked PP \cite{stochastic_geometry2009,BB2010}
%\begin{equation}
$\tilde{\Phi} = \{(x_i,\mvec_{x_i})\},$
%\end{equation}
where $\Phi_p = \{x_i\}$ is the PP of cluster centers  from Def. \ref{def:parentp}, and $\mvec_{x_i}$ is a mark vector, containing all the RVs characterizing the users of the cluster at $x_i$, which are:
%\begin{equation}
%\tilde{\Phi} = \{(x, N_{x,u},N_{x,r}, \uvec_x,\rvec_x,\tilde{\avec}_x,\vvec_x, \hvec_x)\}, \label{eq:adminnet}%\text{where:}
%\end{equation}
%where:
\begin{figure} [!t]
	\centering
	%\begin{tikzpicture}
	%\begin{axis}[grid=none,ticks=none,x=0.3cm, y=0.3cm,ymin = -11, ymax = 11,xmin=-11,xmax=11,
	%legend entries = {Unclustered reqs.,Unclustered users w/files,Clustered. reqs., 
	%Clustered users w/files},
	%legend style = {font = \small},legend pos = south west]
	%\addplot[only marks, cyan,mark = triangle*,mark size =1.5pt] table {Dats/puntos4.dat};
	%\addplot[only marks, green,mark = diamond*,mark size =1.5pt]  table {Dats/puntos2.dat};
	%\addplot[only marks, red,mark = square*,mark size =1pt] table {Dats/puntos3.dat};
	%\addplot[only marks, blue,mark = *,mark size =1.5pt] table {Dats/puntos1.dat};
	%\draw[fill=cyan!30] (axis cs: 9.958 , 6.4302) circle [radius= 1.98];
	%\draw[fill=cyan!30] (axis cs: 5.5413 , 6.9031) circle [radius= 1.98];
	%\draw[fill=cyan!30] (axis cs: 8.7335 , -3.8737) circle [radius= 1.98];
	%\draw[fill=cyan!30] (axis cs: -0.094239 , -5.4117) circle [radius=1.98];
	%\draw[fill=cyan!30] (axis cs: -9.038 , 9.6551) circle [radius= 1.98];
	%\draw[fill=cyan!30] (axis cs: -0.93964 , 5.9342) circle [radius= 1.98];
	%\draw[fill=cyan!30] (axis cs: -2.8166 , 0.77309) circle [radius= 1.98];
	%\draw[fill=cyan!30] (axis cs: 3.7015 , -9.6484) circle [radius= 1.98];
	%\draw[fill=cyan!30] (axis cs: 5.2363 , 2.2483) circle [radius= 1.98];
	%\draw[fill=cyan!30] (axis cs: -9.1784 , -3.6853) circle [radius= 1.98];
	%\draw[fill=cyan!30] (axis cs: -8.8176 , 3.3584) circle [radius= 1.98];
	%%\draw[black!15] (axis description cs:0,0) rectangle (axis description cs:1,1);
	%\end{axis}
	%\end{tikzpicture}
	\begin{tikzpicture}[scale=1]
	%circlulos de clusters
	\foreach \x / \y in {3.15/1, 2/3, 5.8/5.42, 1.06/5.72, 5.78/3.2, 3.76/4, 5.5/1.1, 1.05/1.2, 3.1/6}{
		\draw[fill=orange!30] (\x,\y) circle [radius= .95];
	};
	% caja externa
	\draw[very thick] (0,0) rectangle(7,7);
	% usuarios que comparten
	\draw plot [only marks, mark=*, mark options={color=red,fill=red, scale=.8}]file {puntoscach.dat};
	%usuarios que piden
	\draw plot [only marks, mark=triangle*, mark options={color=blue,fill=blue, scale=1.2}]file {puntosreq.dat};
	%leyenda
	\begin{scope}[shift={(.3,.3)}] 
	\draw[fill=yellow!20,thick] (-.2,-.2) rectangle (3.5,1.3);
	\draw[] (0,0) -- 
	plot[only marks, mark=square*, mark options={fill=orange!30, scale=1.2}] (0,1)
	node[right,font=\footnotesize]{Clusters};
	\draw[] (0,0) -- 
	plot[only marks, mark=triangle*, mark options={color=blue,fill=blue, scale=1.2}] (0,.5)
	node[right,font=\footnotesize]{Users caching files};
	\draw[] (0,0) -- 
	plot[mark=*, mark options={color=red,fill=red, scale=.6},only marks] (0,0.05)
	node[right,font=\footnotesize]{Users sharing files};
	\end{scope}
	\end{tikzpicture}
	\caption{Representation of the network with clusters. The users in the clusters, form the cluster PP $\Phi_c$ according to (\ref{eq:phic1}), in which D2D  takes place.} 
	\label{fig:netrep}
\end{figure}
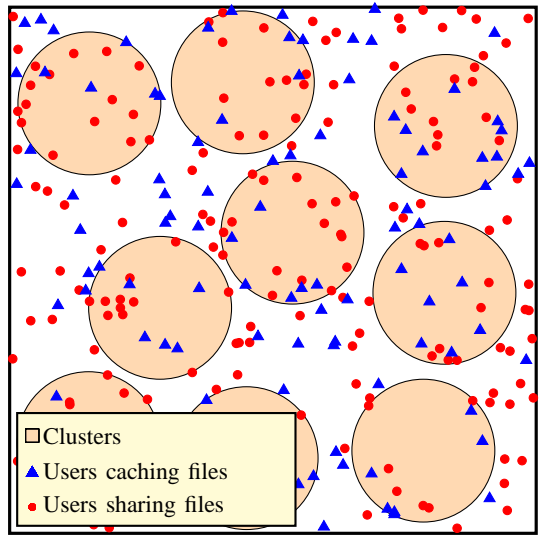 
\begin{itemize}
%\item $\Phi_p = \{x\}$ is an admissible parent process, a hard-core point process of intensity $\lambda_p$ giving the cluster centers.
%As described in section \ref{sec:spatialmodel}, the hard-core parent process $\Phi_p$ is obtained by a position-dependent thinning of this process.
% $\{Z_x\}$ are $U(0,1)$ RVs which are used for the thinning if a type II Matérn process is used as a parent process (unused  for type I processes).
\item $N_{x,u}$: the number of users which cache videos in the cluster centered at $x$. They are Poisson RVs of mean $\lambda_u \pi R_c^2$. $\uvec_x = (S_{x,1}, \ldots, S_{x,N_{x,u}})$ is the vector positions of these users  relative to $x$, which, conditioned on $N_{x,u}$, are i.i.d. uniform RVs on the cluster (Remark \ref{req:equivmod}).
\item $N_{x,r}$: the number of users requesting videos within the cluster centered at $x$. They are Poisson RVs of mean $\lambda_r \pi R_c^2$. $\rvec_x = (D_{x,1}, \ldots, D_{x,N_{x,r}})$ is the vector of positions of these users relative to $x$, which, conditioned on $N_{x,r}$, are i.i.d. uniform RVs on the cluster (Rem. \ref{req:equivmod}).
\item $\tilde{\avec}_x= \{\avec_{x,1}, \ldots, \avec_{x,N_{x,u}}\}$ are the videos which are stored in the users, such that $S_{x,i}$ stores $\avec_{x,i} = (A_{x,i,1},\ldots,A_{x,i,M})$.  They are selected independently according to $p_A$ as indicated before. $\vvec_x = \{V_{x,1}, \ldots, V_{x,N_{x,r}}\}$ are the requested videos such that $D_{x,i}$ requests $V_{x,i}$. They are selected independently according to $p_V$, as mentioned earlier.
\item $\hvec_x$ is a family of  independent power fading coefficients between the users inside the cluster and towards users in other clusters.
\end{itemize}
When clear from context the subscript $x$ in $\mvec_x$ is omitted. The dependence between the variables in the mark vector is characterized by their joint distribution $F_{\mvec_x}$, which, from previous assumptions, factors as:
% The caching and requesting users are distributed as independent Poisson PPs in the clusters (see Remark~\ref{req:equivmod}), and considering the independent requests and caching strategy, this distribution factors as:
\begin{IEEEeqnarray}{rCl}
F_{\mvec_x} % &= F_{N_{x,u},N_{x,r},\uvec_x,\rvec_x,\tilde{\avec}_x,\vvec_x,\hvec_x} \\ %= F_{\uvec_x,\rvec_x,\tilde{\avec}_x,\vvec_x,\hvec_x|N_{x,u},N_{x,r}}F_{N_{x,u},N_{x,r}} \\
&=& F_{\hvec_x|\uvec_x,\rvec_x,N_{x,u},N_{x,r}} F_{\uvec_x|N_{x,u}}F_{\tilde{\avec}_x|N_{x,u}} F_{\rvec_x|N_{x,u}} F_{\vvec_x|N_{x,r}}F_{N_{x,u}}F_{N_{x,r}},\\
&=&{N_{x,u}} F_{N_{x,r}} \left(\prod_{i=1}^{n_r}F_{V_{x,i} |N_{x,r}=n_r} F_{D_{x,i} |N_{x,r}=n_r} \right)
\left[ \prod_{i=1}^{n_u} F_{S_{x,i} |N_{x,u}=n_u} \left(\prod_{j = 1}^M F_{A_{x,i,j} |N_{x,u}=n_u}\right)\right] \nonumber \\ &&\times F_{\hvec|\uvec_{x},\rvec_x,N_{x,u},N_{x,r}}. \label{eq:Ffact1}
\end{IEEEeqnarray}
For shortness, unless mandatory like in the last step, we have not included the point where the distributions are evaluated. For example, $F_{\rvec_x|N_{r,x}} \equiv F_{\rvec_x|N_{r,x}=n_{r,x}}(\mathbf{d}_x)$. 

There are several hard-core PPs which can be used as a parent PP $\Phi_p$. Here we consider two possible examples. One is the  type II Matérn hard-core model~\cite{stochastic_geometry2009}, which is obtained through a position-dependent thinning of a Poisson PP $\Phi$ of intensity $\lambda$;
a uniform RV is drawn for each point of $\Phi$ and, for each pair of points which are separated by less than $\delta$, only the one with the smallest RV is kept. This leads to a cluster density $\frac{(1-e^{-\lambda \pi \delta^2})}{\pi \delta^2}$.
We also consider the translated-grid PP \cite{stochastic_geometry2009}, which gives a regular square grid. It is obtained by using two independent uniform RVs in $[0,\delta)$, $U_1$ and $U_2$, and by considering the grid formed by the pairs $(m \delta+U_1,n\delta+U_2)$, for all integers $m$, $n$. This gives a density of clusters of $\delta^{-2}$.
%
%Two examples are Matérn's hard-core models \cite{stochastic_geometry2009}, obtained from a Poisson PP $\Phi$ of intensity $\lambda$, in which certain points are deleted through a position-dependent thinning. For example, in type II PPs, a uniform RV is drawn for each point of $\Phi$ and, for each pair of points which are separated by less than $\delta$, only the one with the smallest uniform RV remains. Another example is the translated-grid PP \cite{stochastic_geometry2009}, which gives a regular square grid. It is obtained by using two independent uniform RVs in $[0,\delta)$, $U_1$ and $U_2$, and by considering the grid formed by the pairs $(m \delta+U_1,n\delta+U_2)$, for all integers $m$, $n$. With these rules, parent processes which guarantee a minimum clearance of $\delta \geq 2R_c$ are obtained. The density of these PPs is:
%\begin{IEEEeqnarray}{rCl}
%%\lambda_p &= \lambda e^{-\lambda \pi \delta^2} \hspace{15mm} &\text{Matérn Type I}, \label{eq:laHC1}\\
%\lambda_p &= &\frac{(1-e^{-\lambda \pi \delta^2})}{\pi \delta^2} \hspace{4mm}\text{Matérn Type II,} \label{eq:laHC2}\\
%\lambda_p &= &\frac{1}{\delta^2} \hspace{20mm} 	\text{Translated grid}.\label{eq:laTG1}
%\end{IEEEeqnarray}
%The density of these PPs is: $\frac{(1-e^{-\lambda \pi \delta^2})}{\pi \delta^2}$ for Matérn type II, and $\frac{1}{\delta^2}$ for the translated grid.

%Other processes may be constructed via usual operations on point processes, like thinning \cite{stochastic_geometry2009}.
%	\begin{IEEEqnarray}
%\vspace{-3mm}
\subsection{In-cluster Communication and Admissible protocols} \label{sec:inclust}

Given a realization of $\tilde{\Phi}$, in each cluster, a memoryless network is defined, where users who cache videos are sources, which have a  subset of all possible messages, and requesting users are receivers, requesting  one of the possible messages each. To conserve the network symmetry, keep a simple structure and not require long-range coordination, we assume: 
\begin{itemize}
\item Transmitters may have different degrees of CSI pertaining only to their own cluster, that is, some knowledge about the cluster, which is contained in its mark vector, $\mvec_x$.
\item Clusters are uncoordinated, interference between them is treated as noise and there is no interaction between them to reduce interference. 
\item Transmissions in the network take place at a rate $R$  in a block, in which all the clusters attempt to serve some or all of the requests inside at the same time, and an average-power constraint of $P$ is imposed on each user.
\end{itemize}

With the above assumptions, we can focus on a single cluster to describe the behavior of any cluster in the network. 

%Let us suppose that conditioned on $\tilde{\Phi}$ we pick a cluster $(x,\mvec_x)$ which contains $n_u$ users with videos and $n_r$ requests for which the video is available in the cluster.  The cluster uses the coding scheme $\zeta(\mvec_x)$ and this defines an achievable rate region $\mathcal{R}(\mvec_x, \tilde{\Phi}) \subset \R_{+}^{n_r}$ of the rates that are achievable from the $n_u$ transmit nodes to the $n_r$ receiver nodes. This is not the region in which all the users decode successfully, but an extended region, which is the union of the regions in which at least one of the users can decode. Then, we may define a vector of information rates $\mathbf{r} = (R, \ldots, R) \in \R_{+}^{n_r}$, in which the $i$-th component is the rate attempted  through the coding scheme towards receiver $i$. If $\mathbf{r} \in \mathcal{R}(\mvec_x, \tilde{\Phi})$ then all the users will decode successfully, while if this does not happen, then a subset of the users will be able to decode. 
%Due to the symmetry, the asymptotic error probability of each user (after averaging over $\tilde{\Phi}$) will be the same. %\vspace{1mm}

\begin{definition}[In-cluster communication strategy] \label{def:coordstrat}
An in-cluster communication  strategy is given by any coding scheme that guarantees  an achievable rate region $\mathcal{R}(\mvec_x, \tilde{\Phi}) \subset \R_{+}^{n_r}$ for the involved cluster  with $n_u$ transmit nodes to the $n_r$ receiver nodes, where a symmetric rate $R$ is attempted to all users.
\end{definition}
%Notice that most practical communication schemes fit the definition of an admissible strategy. 
\begin{definition}[Served request]
A request from the $i$-th user at the cluster centered at $x$ is said to be served whenever:
\begin{itemize}
\item The video is available in the cluster, that is, there is a match for this user, an event which writes as:
%\begin{equation}
$\M_{x,i} = \bigcup_{j = 1}^{N_{x,u}} \{V_{x,i} \in \avec_{x,j}\}.$ % \label{eq:matchev}
%\end{equation}
\item The transmission is scheduled during the block, that is, the user with the match is scheduled to receive a transmission from one or more users with the video.
\item The $i$-th  transmission rate $R$ belongs to the rate-region $\mathcal{R}(\mvec_x, \tilde{\Phi})$ induced by the strategy.
\end{itemize}
\end{definition}
The probability of having a match is the same for all the users in the network, i.e., $\prob(\M_{x,i}) \triangleq p_{\M}$. However, these events are correlated between users because they use the same cache.
%However, within a cluster, these events are correlated for different users,  because they search for the videos in the same distributed cache.
%We now define the family of admissible protocols:
%\vspace{1mm}
\begin{definition}[Admissible protocol] \label{def:adprot}
An admissible protocol is any pair $(\Phi_p, \mathcal{F})$, where $\Phi_p$ is an admissible parent PPs, which defines a clustered network $\tilde{\Phi}$ and $\mathcal{F}$ is an in-cluster communication  strategy as in Def. \ref{def:coordstrat}.
\end{definition}

%\vspace{-2mm}
\subsection{Performance metrics and trade-offs} \label{sec:metrics}
For every admissible protocol $(\Phi_p, \mathcal{F})$ and given a compact set $K \subset \R^2$ we define $N_s(K,\Phi_p,\mathcal{F},R)$ as the number of served requests in $K$ during a transmission block:
\begin{equation}
N_s(K,\Phi_p,\mathcal{F},R) \triangleq \sum_{x \in \Phi_p} \sum_{i= 1}^{N_{x,r}} \bigone_{\K_{x,i}} \bigone_{\Sset_{x,i}}, \label{eq:Ns1}
\end{equation}
where $\Sset_{x,i} = \{ \text{Req. of user $i$ in $(x,\mvec_x)$ is served}\}$ and
%\begin{equation}
$\K_{x,i} = \{x + D_{x,i} \in K \}$. % \label{eq:evKDxi}.
%\end{equation}
Also, we define $N_{sc}(\Phi_p,\mathcal{F},R,x)$ as the number of served requests in a cluster centered at $x$:
\begin{equation}
N_{sc}(\Phi_p,\mathcal{F},R,x) \triangleq \sum_{i = 1}^{N_{x,r}} \bigone_{\Sset_{x,i}}.
\end{equation}
\begin{lemma} \label{teo:Avreq}
Given a compact set $K \subset \R^2$, the average number of served requests in $K$ is:
\begin{equation}
\Ex \left[ N_s(K,\Phi_p,\mathcal{F},R)\right]  =\lambda_p |K|\Ex^0[N_{sc}(\Phi_p,\mathcal{F},R,0)],\label{eq:LogGlobRel}
\end{equation}
where $|K|$ is the area of $K$. $\Ex^0$ is the expectation with respect to the Palm distribution of the PP with a cluster at the origin, a conditional distribution of the realizations of the PP with a cluster at the origin.
 The term
$\Ex^0[N_{sc}(\Phi_p,\mathcal{F},R,0)]$ is the average number of users served in any cluster in the network.
\end{lemma}
\begin{IEEEproof} Please see Appendix \ref{proof:Avreq}. \end{IEEEproof}
We next define the main metrics under study. 

\begin{definition}[Local metric] \label{def:locmet}
The ratio of mean served requests per cluster is:
\begin{equation}
T_L(\mathcal{F},R) = \frac{\Ex^0[N_{sc}(\Phi_p,\mathcal{F},R,0)]}{\Ex^0[N_{0,r}]}, \label{eq:TC1}
\end{equation}
where $\Ex^0[N_{0,r}] = \lambda_r \pi R_c^2$ is the average number of requests within any cluster of the network. This ratio indicates how many requests are served on average in any cluster of the network, relative to the average number of requests per cluster.
\end{definition}
Linking this metric with (\ref{eq:LogGlobRel}) we define a global metric:
\begin{definition}[Global metric] \label{def:globmet} Chosen a compact set $K$, the ratio of mean served requests is:
\begin{align}
T_G(\mathcal{F},R)&= \frac{\Ex \left[ N_s(K,\Phi_p,\mathcal{F},R)\right]}{\Ex[N_r(K)]} \\
 &= \lambda_p |\B(0,R_c)| T_L(\mathcal{F},R), 
 \label{eq:TG1}
\end{align}
where $\Ex[N_r(K)] =\lambda_r |K|$ is the average number of requests in the set $K$.
Due to the stationarity of the PP, this does not depend on $K$ and it can be interpreted as the spatial density of served requests, normalized by the  density of requests $\lambda_r$.
\end{definition}
These metrics are determined by many factors, such as the caching policy and video request statistics, the attempted rate $R$,  the cluster radius, and the in-cluster communication  strategy.
%There is a trade-off behavior in these metrics. 
The local metric always benefits from a reduction in the cluster density because as $\lambda_p \rightarrow 0$ the interference decreases on average. If the cluster radius $R_c$ is fixed and  $\lambda_p$ is diminished, the local metric will benefit, but if the density becomes too small, the global metric will eventually have to decrease. This means  there is a trade-off between the metrics.

\begin{definition}[Average rate]
Given an admissible protocol $(\Phi_p, \mathcal{F})$, a cluster at the origin	 has an average rate:
\begin{equation}
\bar{R}(\mathcal{F},R)=R \ \Ex^0\left[ \sum_{i=1}^{N_{0,r}} \frac{\bigone_{\Sset_{0,i}}}{ \sum_{j=1}^{N_{0,r}} \bigone_{\M_{0,j} \cap \mathcal{P}_{0,j}} }  \right], \label{eq:defRbar}
\end{equation}
where $\mathcal{P}_{0,j}$ is the event that a transmission to user $j$ is scheduled in the transmission block.
%This involves averaging over all the transmissions in a cluster for each realization of $\tilde{\Phi}$ and also to $\tilde{\Phi}$.
\end{definition}
Considering the metrics in Defs. \ref{def:locmet} and \ref{def:globmet} with an average-rate constraint, which models requirements in terms of delay and link-reliability, we define the following trade-off regions: 
\begin{definition}[Trade-	off regions] \label{def:tradeoffs}
\begin{itemize}
\item \emph{Global-metric trade-off  region:} a pair $(r,t)$ of average rate and global metric is said to be achievable if there exists an admissible protocol $(\breve{\Phi}_p,\breve{\mathcal{F}})$ with rate $R$ and density $\lambda_p$ satisfying:
\begin{equation}
\begin{cases}
T_G(\breve{\mathcal{F}},R) \geq  t,\\
\bar{R}(\breve{\mathcal{F}},R) \geq r. \label{eq:TgRtoff}
\end{cases}
\end{equation}
The set of all achievable pairs $(r,t)$ is the global-metric trade-off  region. 

\item \emph{Local-metric trade-off  region:} a tuple $(r,t,\lambda_l)$ of average rate, local metric and parent density is said to be achievable if there exists an admissible protocol $(\breve{\Phi}_p,\breve{\mathcal{F}})$ with rate $R$ and density $\lambda_p$ satisfying:
\begin{equation}
\begin{cases}
T_L(\breve{\Phi}_p,\breve{\mathcal{F}},R)  \geq  t, \\
\bar{R}(\breve{\Phi}_p,\breve{\mathcal{F}},R) \geq r,\\
\lambda_p(R_c, \delta) \geq \lambda_l. \label{eq:TcRtoff}
\end{cases}
\end{equation}
The set of all achievable tuples $(r,t,\lambda_l)$ is the  local-metric trade-off  region.

\item \emph{Local-global trade-off  region:} given an attempted rate $R$, a pair $(t_g,t_c)$ of global and local metrics is said to be achievable if there exists an admissible protocol $(\breve{\Phi}_p,\breve{\mathcal{F}})$ with rate $R$ and density $\lambda_p$ satisfying:
\begin{equation}
\begin{cases}
T_G(\breve{\Phi}_p,\breve{\mathcal{F}},R) \geq  t_g,\\
T_L(\breve{\Phi}_p,\breve{\mathcal{F}},R) \geq  t_c. \label{eq:TgTctoff}
\end{cases}
\end{equation}
The set of all achievable pairs $(t_g,t_c)$ is the  local-global trade-off  region.
\end{itemize}
\end{definition}

The first region refers to the maximum fraction of users which receive videos successfully globally, subject to an average rate constraint. This metric is limited by the fraction of the users of the network that are clustered, because unclustered users cannot exchange videos. A PP $\breve{\Phi}_p$ should be chosen such that the network is almost fully covered by clusters, and a strategy  $\breve{\mathcal{F}}$ such that  all the requests in a cluster can be served ($T_L \approx 1$), while fulfilling the rate constraint. 
The second region refers to the fraction of the users inside a cluster that receive videos successfully. In this case, both a rate constraint and a certain density of clusters are required. 
Otherwise, we could set $\lambda_p \approx 0$ and a achieve a large level of service at the typical cluster, but there would be almost no other cluster in the network. Since this region is defined by what happens in a cluster, we could have $T_L \approx 1$, for any clustering PP.
The third region refers to maximizing one the metrics, with a constraint on the other one, balancing the global and local benefits of D2D.

We cannot find the optimal protocol in terms of each trade-off region. However, analyzing a specific protocol  will yield inner regions and thus yield insights on the performance attainable through D2D. 

The following Lemma provides a straightforward bound for the local and global metrics (proof found in Appendix \ref{proof:toffbou}). %, and hence, to the trade-off regions.  \vspace{1mm}

\begin{lemma} \label{teo:toffbou}
Given a protocol $(\Phi_p, \mathcal{F})$, we have
%\begin{equation}
$T_L(\mathcal{F},R) \leq p_{\M},$ % \label{eq:bouTC}
%\end{equation}
with equality as $R \rightarrow 0$. This implies for the global metric that
%\begin{equation}
$T_G(\mathcal{F},R) \leq  \lambda_p p_{\M} |\B(0,R_c)|.$  %\label{eq:bouTG}
%\end{equation}
In addition, under the caching scheme described, the probability of a match is:
%\begin{equation}
$p_{\M} =1 - \Ex_{V} \left[ e^{-\lambda_u \pi R_c^2 [1-\left( 1- p_{A}(V)\right)^M ]} \right].$ %\label{eq:pm1}
%\end{equation}
\end{lemma}
%\begin{IEEEproof}
%See Appendix \ref{proof:toffbou}.
%\end{IEEEproof}
This Lemma states that if no constraint is imposed on the transmission quality, the local metric $T_L$ is limited by the probability of finding a match in a cluster $p_{\M}$, and the global metric $T_G$ is limited by this probability and the fraction of clustered users,  which is $\lambda_p |\B(0,R_c)|$.
%This lemma is quite reasonable, since the mean ratio of served requests in a cluster can never exceed the fraction of users which can have a match in the cluster. In addition as $R\rightarrow 0$ the probability of failed transmissions vanishes and only the problem of finding a match remains. The right side of (\ref{eq:bouTG}) is the spatial density of matches in the network. It is clear that if we introduce additional constraints on the average rate or the density of the parent process, the bounds will still be valid, so they will give outer bounds on the regions achievable if we fix the parent process.
%In the numerical results section we shall verify and explain the application of this Lemma to the case of a type II Matérn  process.

\section{Analysis of an in-cluster communication strategy} \label{sec:stratan}

\subsection{Strategy definition} \label{sec:inclusstrat}
In this section we define and analyze an in-cluster communication strategy which, paired with any parent PP, forms an admissible D2D protocol via Def. \ref{def:adprot}. We focus on the cluster at the origin $(0,\mvec_0)$, omitting the subscript $0$ in all its RVs ($N_{0,u} \equiv N_u$, etc.).

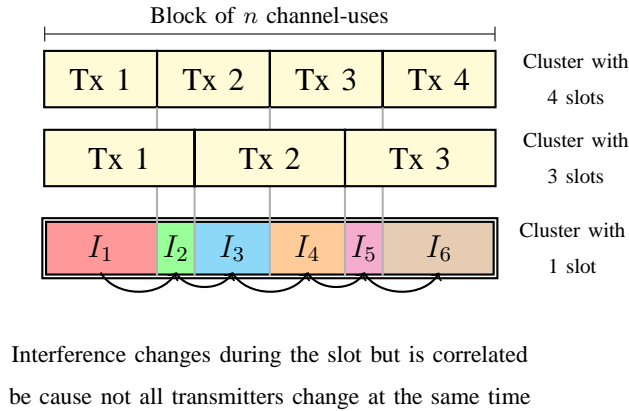
\begin{figure} [!t]
	\centering
	\begin{tikzpicture}[caja/.style = {thick,fill=yellow!20},linea/.style={thick,-,black!30},scale=1.5]
	% Cluster con 4 slots
	\draw [|-|] (0,.65) to  node[above,font= \footnotesize] {Block of $n$ channel-uses} (4,.65) ;
	\draw[caja](0,0)  rectangle node (Tx1a) {Tx 1}(1,.5) 
					  rectangle node(Tx2a){Tx 2}(2,0) 
					  rectangle node(Tx3a){Tx 3}(3,0.5)
					  rectangle node(Tx4a){Tx 4}(4,0) ;
	\node [right = 0.4cm of Tx4a,align=center,font = \scriptsize] (Cfour)  {Cluster with \\4 slots};
	% Cluster con 1 slot
	\fill [fill=red!40] (0,-1.5) rectangle node(Int1) {$I_1$}  (1,-1);
	\fill [fill=green!40] (1,-1.5) rectangle node(Int2) {$I_2$}  (1.333,-1);
	\fill [fill=cyan!40] (1.333,-1.5) rectangle node(Int3) {$I_3$}  (2,-1);
	\fill [fill=orange!40] (2,-1.5) rectangle node(Int4) {$I_4$}  (2.666,-1);
	\fill [fill=magenta!40] (2.666,-1.5) rectangle node(Int5) {$I_5$}  (3,-1);
	\fill [fill=brown!40] (3,-1.5) rectangle node (Int6) {$I_6$}  (4,-1);
	\draw [thick,double] (0,-1.5) rectangle node (clust1match){} (4,-1);
	\node [right = 0.6cm of Int6,align=center,font = \scriptsize] (Cone)  {Cluster with \\1 slot};
	\draw [linea] (1,0) to (1,-1.5);	
	\draw [linea] (1.333,-.2) to (1.333,-1.5);
	\draw [linea] (2,0) to (2,-1.5);
	\draw [linea] (2.666,-.2) to (2.666,-1.5);	
	\draw [linea] (3,0) to (3,-1.5);
	
%	% Cluster con 3 slots
	\draw [caja] (0,-.7) rectangle node (Tx1b) {Tx 1} (1.333,-.2) 
				rectangle node (Tx2b) {Tx 2} (2.666,-.7)
				rectangle node (Tx3b) {Tx 3} (4,-.2);
	\node [right = 0.65cm of Tx3b,align=center,font = \scriptsize] (Cthree)  {Cluster with \\3 slots};
%	
%	%Flechas
	\draw [->,thick] (Int1.south) to [bend right=60] (Int2.south);
	\draw [->,thick] (Int2.south) to [bend right=60] (Int3.south);
	\draw [->,thick] (Int3.south) to [bend right=60] (Int4.south);
	\draw [->,thick] (Int4.south) to [bend right=60] (Int5.south);
	\draw [->,thick] (Int5.south) to [bend right=60] (Int6.south);
	\node [below = 1 of clust1match,font= \footnotesize,align = center] {Interference changes during the slot but is correlated\\ 
	be cause not all transmitters change at the same time} ;
	\end{tikzpicture}
	\caption{Dividing the transmission block in a number of slots equal to the number of matches in a cluster, makes the interference time-correlated and non-stationary, because transmissions among clusters are unsynchronized.} 
	\label{fig:intcorr}
\end{figure}

To consider a simple strategy we assume that:
\begin{itemize}
\item To reduce the interference inside the clusters, a TDMA scheme is employed to divide the transmission block into equal sized slots, in which only one request is served, through a point-to-point transmission  of power $P$. 
\item At most $n_{m,\max}$ slots are defined in each cluster, regardless of the number of matches. 
This does not have any practical implications, since we can choose this such that $F_{N_m}(n_{m,\max}) \approx 1$, i.e., fraction of clusters with dropped requests is negligible\footnote{This requirement is considered for mathematical reasons. In any cluster, the number of slots/matches is always finite, but, since there is an infinite number of clusters, the maximum number of slots over all clusters is unbounded, which is in conflict with the finite (yet long) length of the block.}. 
\end{itemize}
To fully occupy the block with this TDMA scheme, each cluster would split the block into as many matches as it has, and clusters with the same number of matches would have the same number of slots. This assignment is reasonable in terms of theoretical performance, but leads to an interference model which is not tractable, mainly because the transmissions between clusters with a different number of matches is unsynchronized.  Fig. \ref{fig:intcorr}, which focuses on a cluster with one slot, is provided to help understand this. When a slot is over in another cluster, its transmitter is replaced by a new one, which changes a part of the interference. Meanwhile, the other clusters will generate the same interference as before. This results in a time-correlated interference, whose statistics are very involved to model, specially considering the large range of number of slots in a very large network. Also, the distribution of the interference is not the same in all slots of the cluster. This is  inherent  to any wireless system using a similar TDMA scheme.
To overcome this, some degree of regularity is required. Splitting the block in $n_{m,\max}$ slots would solve the problem but lead to an inefficient use of resources, since the block would be mostly unoccupied in all clusters. 
%For this, we propose a strategy which splits the block in a number of slots which is a power of two. 
For this, we propose a strategy in which the block always has a power of two number of slots. 
This allows a different number of slots between clusters, improves the use of resources, and considers interference changes during a slot. 
%The $i$-th set contains the indexes of the transmitters which may potentially serve the $i$-th user's request. 
%Any in-cluster coordination strategy has to choose the users which  transmit the videos to the corresponding destination from these sets.
\begin{definition}[Strategy] \label{def:strat}
In a cluster with $N_m$ matches:
\begin{itemize}
\item The block is split into $W(N_m,\varepsilon)$ slots:
\begin{equation}
W(N_m,\varepsilon)= \begin{cases}
W_L(N_m) & \text{if }\frac{N_m-W_L(N_m)}{W_H(N_m) - W_L(N_m)} < \varepsilon,\\
W_H(N_m) & \text{otherwise,} \label{eq:defLNm} \IEEEeqnarraynumspace
\end{cases}
\end{equation}
where $0 < \varepsilon \leq 1$ is a design parameter, and:
%\begin{alignat}{2}
$$W_H(N_m) = 2^{\lceil\log(N_m)\rceil}, \ \ \ \ \  
W_L(N_m) = 2^{\lfloor\log(N_m)\rfloor},$$
%\end{alignat}
 the powers of two closest to $N_m$ from above and below, respectively.
The RVs $W(N_m,\varepsilon) \equiv W_{x}(N_{m,x},\varepsilon)$, defined for each cluster, are independent like the $\{N_{m,x}\}$.
%If $T_H(N_m)$ slots are used, all the requests with matches can be scheduled for transmissions, while some slots may be left empty. If $T_L(N_m)$ are used, then $N_m - T_L(N_m)$ matches have to be dropped, but all the slots are filled.  
\item For each request with a match, a caching user is selected at random from the set of users who have the video:
%\begin{equation}
$\Aset(V_i,\tilde{\avec}) = \{ j: V_i \in \avec_j, 1\leq j \leq N_u\},$ %\label{eq:setA1} 
%\end{equation}
as a candidate to  serve the request. 
\item If there are $W = W_H(N_m)$ slots, transmissions are scheduled by selecting $N_m$ out of the $W_H(N_m)$, and generating a random permutation of the transmissions in these slots. Otherwise, $N_m-W_L(N_m)$ requests are dropped at random, and the rest are assigned by generating a random permutation of the slots.
\end{itemize}
\end{definition} 
The parameter $\varepsilon$ balances the fraction of time in which the channel is occupied per cluster, with the fraction of the requests that are served. When $\varepsilon = 0$, all the requests are served, maybe leaving some slots unused. When $\varepsilon = 1$ some requests are dropped, but  all the slots are occupied. Also, by dropping requests, a higher success probability per user may be achieved because each transmission gets more channel uses\footnote{We haven't considered the possibility that $N_m > n_{m,\max}$. We could do this, but it would further complicate the exposition without adding any substantial modifications, because both the probability of this event in the typical cluster, and the fraction of the clusters in this condition are negligible.}. Since only one transmitter is chosen at random from $\Aset(V_i,\tilde{\avec})$ to serve a request, for each user $1\leq i \leq N_r$ in a cluster, we define an  RV $C_i$ indicating which user transmits:
\begin{equation}
C_i|_{V_i,\tilde{\avec}} \sim U(\Aset(V_i,\tilde{\avec})) \ \ \ \text{if }\Aset(V_i,\tilde{\avec}) \neq \emptyset, \label{eq:distC}
\end{equation}
where $U(\cdot)$ denotes a uniform RV over the set, and $C_i = 0$ if $\Aset(V_i,\tilde{\avec}) = \emptyset$.

\subsection{Interference characterization and achievable rates}
We now analyze the interference generated by our scheme, to define the achievable rates of the users.  We assume that $\tilde{\Phi}$ is fixed, and also, that transmissions have been scheduled in each cluster, according to Def. \ref{def:strat}.   We consider a fixed point in the cluster at the origin, during one of the slots, which we call the \emph{slot under study}. Notice that the interference seen by a user depends on the total number of slots in its own cluster. %, which implies that selecting this also impacts the interference seen by a user. This is opposed to usual scenarios, in which the interference seen by a user is constant and does not depend  on what the user does.

\begin{figure} [!t]
\centering
\begin{tikzpicture}[caja/.style = {very thick,fill=yellow!20},linea/.style={thick,-,black!30},scale=1.5]
% Cluster con 4 slots
\draw[caja](0,0)  rectangle node (Tx1a) {Tx 4}(1,.5) 
	rectangle node(Tx2a){Tx 5}(2,0) 
	rectangle node(Tx3a){Tx 6}(3,0.5)
	rectangle node(Tx4a){Tx 7}(4,0) ;
\node [right = 0.4cm of Tx4a,align=center,font = \scriptsize] (Cfour)  {Cluster with \\ $4n_1$ slots};
% Cluster con 1 slot
\fill [fill=red!40] (0,-1.5) rectangle node(Int1) {$I_1$}  (1,-1);
\fill [fill=green!40] (1,-1.5) rectangle node(Int2) {$I_2$}  (2,-1);
\fill [fill=cyan!40] (2,-1.5) rectangle node(Int3) {$I_3$}  (3,-1);
\fill [fill=orange!40] (3,-1.5) rectangle node(Int4) {$I_4$}  (4,-1);
\node [right = 0.5cm of Int4,align=center,font = \scriptsize] (Cone)  {A single slot in a cluster \\with $n_1$ slots};
\draw [thick,double] (0,-1.5) rectangle node (clust1match){} (4,-1);
\draw [linea] (1,0) to (1,-1.5);	
\draw [linea] (2,0) to (2,-1.5);
\draw [linea] (3,0) to (3,-1.5);
% Cluster con 3 slots
\draw [caja] (0,-.7) rectangle node (Tx1b) {Tx 2} (2,-.2) 
rectangle node (Tx2b) {Tx 3} (4,-.7);
\node [right = 1.2cm of Tx2b,align=center,font = \scriptsize] (Cfour)  {Cluster with \\ $2n_1$ slots};

%% Cluster con menos slots
\draw[caja] (0,-1.8) rectangle node (Tx0a) {Tx 1}(4.4,-2.3) ;
\node [right = 2.9cm of Tx0a,align=center,font = \scriptsize] (Cthree)  {Cluster with at \\most $n_1$ slots};

\end{tikzpicture}
\caption{We focus on a single slot in a cluster with $n_1$ slots (double line).
	Clusters with more than $n_1$ slots cause time variations in the interference, because different transmitters are active in each slot. Clusters with at most $n_1$ slots cause a constant interference, because only one transmitter is  active.}
\label{fig:syncint}
\end{figure}
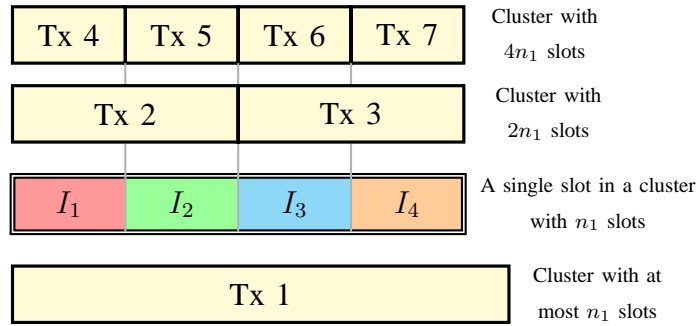

Assume that the cluster at the origin has $n_1$ slots, $n_1$ being a power of two, as indicated in Def. \ref{def:strat}. Fig. \ref{fig:syncint} shows how the interference behaves during the slot under study:
\begin{itemize}
\item Clusters with at most $n_1$ slots generate a constant interference power because only one transmitter is active. 
\item Clusters with more than $n_1$ slots, say $2^k n_1$,  will generate $2^k$ interference powers during the slot. 
\end{itemize} 
If the maximum number of slots in a cluster is $\Delta \triangleq W_H(n_{m,\max})$, there will be $\Delta/n_1$ different interference powers during the slot under study. If $\{\tilde{I}_1(y), \ldots, \tilde{I}_{\Delta/n_1}(y)\}$ are these powers,  and we consider a long block-length, a transmission from $x$ to $y$ could achieve a rate $R$:
\begin{equation}
R < \frac{1}{\Delta} \sum_{i = 1}^{\Delta/n_1} \C\left(\text{SIR}_i\right), \label{eq:achrate}
\end{equation}
where $\text{SIR}_i = |g_{xy}|^2 / \tilde{I}_i(y) .$
\begin{figure*}[b!] 
	\footnotesize
	\hrulefill
	\vspace{-2mm}
	\setcounter{MYtempeqncnt}{\value{equation}} % guardo el valor
	\setcounter{equation}{21}
	\begin{equation}
	T_L(\mathcal{F}^*,R) = \frac{\Ex^0 \left[ \left(N_m  \bigone_{\{W(N_m) = W_H(N_m)\}} +  W_L(N_m) \bigone_{\{W(N_m) = W_L(N_m)\}}\right) \bar{F}_{|g_{S,D}|^2|S,D}\left(I(D,W(N_m)) (2^{W(N_m)R}-1)\right)\right]}{\Ex^0[N_r]}.\! \! \!   \label{eq:LocalM1}
	\end{equation}	
\end{figure*}
Notice that SIR changes are caused by the interference, while the source-destination channel is the same.
%This rate is obtained because there are a total of $\Delta$ interference power during a block, and a slot will see $\Delta/n_1$ of these. 
%This rate does not depend on the rates achievable by other users, which means that the rate region $\mathcal{R}$ defined in Sec. \ref{sec:inclust} is a hyperrectangle.
The complexity of (\ref{eq:achrate}) precludes finding the probability of a failed transmission, i.e. the probability that (\ref{eq:achrate}) does not happen, so we develop a lower bound on the achievable rate. To do this, we consider that all the clusters with at most $n_1$ slots will generate a constant interference, which we can add to define $I_b(n_1,y)$:
\setcounter{equation}{\value{MYtempeqncnt}} % restore previous value
\begin{equation}
I_b(n_1,y)=\sum_{j=0}^{\log n_1} I_j(y), \label{eq:Ib}
\end{equation}
where $I_j(y)$ is the aggregate interference at $y$ from the clusters with $2^j$ slots during the slot under study. Clusters with more than $n_1$ slots, say $2^kn_1$, generate $2^k$ interference values; we index these $2^k$ values using a binary expansion, as $I_k(u_1,\ldots,u_k,y)$, with $(u_1,\ldots,u_k) \in \{1,2\}^k$, for any $1 \leq k \leq \log (\Delta/n_1)$. This indexing is only used because it is convenient for the proof of Lemma \ref{teo:bourate}. We do not need to specify and order in which these interferences appear during the slot under study; we are only interested in the time average of these values:
\begin{equation}
\bar{I}_k(n_1,y) = \frac{1}{2^k} \sum_{(u_1,\ldots,u_k) \in \{0,1\}^k} I_k(u_1,\ldots,u_k,y). \label{eq:Iav}
\end{equation}
That is, $\bar{I}_k(n_1,y)$ is the time-average of the interference coming from clusters with $2^kn_1$ slots, seen at $y$ during the slot under study. With these definitions we introduce the following:
\begin{lemma}[Achievable rate] \label{teo:bourate}
The rate (\ref{eq:achrate}) of any slot in a cluster with a total of $n_1$ slots can be lower bounded by:
\begin{equation}
R_a(n_1,x,y)=\frac{1}{n_1}\C\left(\frac{P|g_{xy}|^2}{I_b(n_1,y) +\sum_{ i = 1}^{\log (\Delta/n1)}  \bar{I}_k(n_1,y)}\right),\label{eq:rach2}
\end{equation}
where $I_b$ comes from (\ref{eq:Ib}) and $\bar{I}_k(n_1,y)$ is given by (\ref{eq:Iav}).
%$I_b(n_1,y)$ is the total interference generated by clusters with at most $n_1$ slots:
%\begin{equation}
%I_b(n_1,y)=\sum_{j=0}^{\log n_1} I_j(y), \label{eq:Ib}
%\end{equation}
%where $I_j(y)$ is the interference generated by the clusters with $2^j$ slots. In addition, $\bar{I}_k(n_1,y)$ is the time average of the powers observed coming from clusters with $2^kn_1$ slots:
%\begin{equation}
%\bar{I}_k(n_1,y) = \frac{1}{2^k} \sum_{(u_1,\ldots,u_k) \in \{1,2\}^k} I_k(u_1,\ldots,u_k,y). \label{eq:Iav}
%\end{equation}
\end{lemma}
\begin{IEEEproof}
Details can be found in Appendix \ref{proof:bourate}.
\end{IEEEproof}

This means that any rate $R < R_a$ is achievable, where $R_a$ is a time-sharing, point-to-point rate with a constant interference, made up of the original interferences added together, with components that change over the slot being time-averaged. We now write the expression of the interference in (\ref{eq:rach2}) as seen from the slot under study.  If we focus on an interfering cluster with $2^kn_1$ slots, we see that the same transmitter can be active in more than one slot, or some slots may be empty, so we cannot assume that there will be $2^k$ different or non-null interference powers. Since we have time-averaged the interference, we are only interested in how many times each user transmits, and not in the transmission order. By assuming that all the slots are occupied, which gives a worst case, the interference in (\ref{eq:rach2}) at a point $d$ of the cluster at the origin with $n_1$ slots, during a slot, is:
\begin{multline}
	I(d,n_1) =  \sum_{x\in\Phi\setminus\{0\} } P\left\{ \bigone_{\{W_x \leq n_1\}} |h_{xd}|^2 l(x+S_x,d)\phantom{\sum_{i =1 + 1}}\right. \\\left.
	+\sum_{i =\log(n_1)+1}^{\log \Delta}\bigone_{\{W_x = 2^i\}} \frac{n_1}{2^i} \sum_{j=1}^{2^i/n_1} B_{x,j,i} |h_{x+S_{x,j},d}|^2 l(x+S_{x,j},d)
	\right\}. \label{eq:intDL1}
	\end{multline}
The terms with $\bigone_{\{W_x \leq n_1\}}$ represent the interference from users in clusters with at most $n_1$ slots. The terms with $\bigone_{\{W_x = 2^i\}}$ represent the interference from a cluster with $2^i$ slots. The RVs $B_{x,j,i}\geq 0 $ indicate how many times each transmitter is active during slot. Since we take that all the slots are occupied, we have:
\begin{equation}
\sum_{j=1}^{2^i/n_1} B_{x,j,i} = 2^i/n_1. \label{eq:condBxij}
\end{equation}
\stepcounter{equation}
If in a cluster $B_{x,j,i}= 1$, for all $j$, then every slot that took place during the slot under study is used by a different transmitter; if only one satisfies $B_{x,j,i}= 2^i/n_1$ and the rest are zero, all the slots are used by the same transmitter. % The joint distribution of these variables, inherited by the in-cluster coordination strategy, will not be required so we won't specify it. 
\begin{remark}
Actually, there is an abuse of notation in (\ref{eq:intDL1}). The terms $|h_{x+S_{x,j},d}|^2 l(x+S_{x,j},d)$ do not use the same indexing over $j$ as in the definition of $\tilde{\Phi}$.
In fact, for example, for a cluster with $2^i>n_1$ slots, we cannot guarantee that there will be $2^i/n_1$ different users with videos stored, i.e. that the sum over $j$ is well defined. 
We have obtained an upper bound to the interference seen at $d$ during the slot under study. To do this, we take each cluster, see which transmitters will be active during the slot under study, and count how many times each will transmit. Then, we may add more fictitious transmitters to have $2^i/n_1$ transmitters, and assign them slots such that (\ref{eq:condBxij}) is met. We do not need to consider which of the users actually transmits in each cluster, because we are focused on a single slot of the cluster at the origin, and because, conditioned on the cluster centers, the fading and user positions are  independent. %The distribution of the variables $\{B_{x,j,i}\}$ will not be required, as we shall see.
\end{remark}

\subsection{Performance metric analysis}
We now evaluate the metrics from Def. \ref{def:tradeoffs} for the strategy given by Def \ref{def:strat}, which for simplicity we denote by $\mathcal{F}^*$. For a compact set $K$ and a protocol $(\Phi_p,\mathcal{F}^*)$, the average number of served requests  is given by (\ref{eq:Ns1}). The event of a served requests $\Sset_{x,i}$ can be written as:
%\begin{equation}
%N_s(K,\Phi_p,\mathcal{F}^*) =  \sum_{x \in \Phi_p} \sum_{i=1}^{N_{x,r}} \bigone_{\K_{x,i}} ,  \label{eq:cntnum}
%\end{equation}
%\begin{equation}
$\Sset_{x,i} = \M_{x,i} \cap \T_{x,i},$ %\label{eq:cntnum}
%\end{equation}
where $\M_{x,i}$ indicates a match, and $\T_{x,i}$ means  a transmission was scheduled and succeeded ($R < R_a$, $R_a$ from (\ref{eq:rach2})).
\begin{theorem}[Local and global metrics] \label{teo:servicep} Given a parent PP $\Phi_p$, as in Def. \ref{def:parentp}, the ratio of mean served requests per cluster, or local metric $T_L$ (\ref{eq:TC1}), for $\mathcal{F}^*$ is given by (\ref{eq:LocalM1}) at the bottom of the page.
%\begin{figure}[b!]
%\begin{multline}
%\hspace{-1mm}T(\Phi_p,\mathcal{F}^*\hspace{-1mm},R) \hspace{-.5mm}\Ex[N_r] = \Ex\left[ N_m  \bigone\{L(N_m) = T_H(N_m)\}  \Ex^!_0\hspace{-1mm}\left[\bar{F}_{|h_{S,D}|^2}\hspace{-1mm}\left(\hspace{-.5mm}\frac{I(D,L(N_m)) (2^{T_H(N_m)R}-1)  }{l(S,D)}\hspace{-.5mm}\right)\right]\right] +
%\\ \Ex\left[ T_L(N_m)  \bigone\{L(N_m) = T_L(N_m)\} \Ex^!_0\hspace{-1mm}\left[\bar{F}_{|h_{S,D}|^2}\hspace{-1mm}\left(\hspace{-.5mm}\frac{I(D,L(N_m)) (2^{T_L(N_m)R}-1)  }{l(S,D)}\hspace{-.5mm}\right)\right]\right] \label{eq:LocalM1}
%\end{multline}
%\end{figure}[b!]
The spatial density of served requests or global metric (\ref{eq:TG1}) for the  strategy $\mathcal{F}^*$ is $T_G(\mathcal{F}^*\hspace{-1mm},R) = \lambda_p |\B(0,R_c)| T_L(\mathcal{F}^*\hspace{-1mm},R) $. 
The average rate achieved is: % ($ W(N_m)$ is given by (\ref{eq:defLNm})): %\stepcounter{equation}
\begin{equation*}
\!\!\hspace{-1mm}\bar{R}(\mathcal{F}^*\hspace{-1mm},R) \hspace{-.5mm} =\! R\ \Ex^0\!\!\left[\bar{F}_{|g_{S,D}|^2|S,D}\hspace{-1mm}\left(\hspace{-.5mm}I(D,W(N_m)) (2^{W(N_m)R}-1)\hspace{-.5mm}\!\right)\right]\!\!. %\label{eq:AvgRate}
\end{equation*}
\end{theorem}
\begin{IEEEproof}
See Appendix \ref{proof:servicep}.
\end{IEEEproof}
Achieving a high level of service within a cluster may imply that only a few clusters need be created, and hence, globally the effects of D2D may not be significant. Locally, larger clusters imply 
an increased likelihood of matches, but also a larger average transmission distance, and a reduced slot duration, which increases the chances of failed transmissions.
%that users are more likely to have matches, but also increases the average distance between users and reduces the duration of a slot, increasing the chances of failed transmissions.

As we mentioned before, the intra-cluster attenuation model could have different expressions. One case of interest is when it is the same as the inter-cluster model (\ref{eq:interc}), that is, when we have $|g_{xy}|^2 = |h_{xy}|^2 l (x,y)$. %, with $|h_{xy}|^2$ a random power attenuation coefficient and $l(x,y)$ the path loss function. 
Also, it is interesting to consider the specific case in which the fading is Rayleigh, that is,  $|h_{xy}|^2$ follows a unit mean exponential distribution. In this case we have:
\begin{equation}
\Ex^0\left[\bar{F}_{|g_{x,y}|^2}\left(\eta I(d,n)\right)\right] %=\Ex^0\left[\exp\left(-\frac{\eta I(d,n)}{l(x,y)}\right)\right] 
= \mathcal{L}^0_{I(d,n)} \left(\frac{\eta}{l(x,y)}\right),  \label{eq:LTred}
\end{equation}
where $\mathcal{L}^0_{I(d,n)} \left( \eta\right) = \Ex^0\left[e^{-\eta I(d,n)}\right]$ is the LT of  $I(d,n)$ with respect to  $\prob^0$.

The main issue in this case is the distribution of the variables $B_{x,j,i}$. Fortunately, the following Lemma, whose proof is in Appendix \ref{proof:LTbou}, helps us avoid this:
\vspace{1mm}
\begin{lemma} \label{teo:LTbou}
The LT  $\mathcal{L}^0_{I(z,n)}$ of (\ref{eq:intDL1}) for  Rayleigh fading can be lower bounded by setting $B_{x,j,i} = 1$ for all $x,j,k$.  \label{teo:LTbou1}
\end{lemma}
%\begin{IEEEproof}
%See Appendix \ref{proof:LTbou}.
%\end{IEEEproof}
Notice that these results are valid for any PP of cluster centers $\Phi_p$. We now consider that $\Phi_p$ is a type II Matérn hard-core PP and approximate the LT (\ref{eq:LTred}).

\subsection{Approximations and bounds for Matérn type II processes} \label{sec:Mat2Apps}
The main issue to evaluate the metrics for a type II PP is finding the reduced LT of the interference (\ref{eq:LTred}). This PP has been used  mostly to model networks using carrier-sense multiple access schemes\cite{BB2010,BaccelliDense2007}, and even the most simple transforms are not known in closed form. For this reason, we approximate the PP by a more tractable one, following the approach in \cite{Haenggi_MeanInt2011}. In our setting this equates to considering the cluster centers, as seen from the cluster at the origin, to be distributed as a non-homogeneous Poisson PP of intensity:
$\lambda_p \bigone_{\{||x|| > \delta \}}$, where $\lambda_p$ is the intensity of the original hard-core PP (Section \ref{sec:userdesc}). 
If we use Lemma \ref{teo:LTbou1} to bound the true interference (\ref{eq:intDL1}), and approximate the Matérn PP  in this way, we have the following approximate interference:
\begin{equation}
I(d,n) \approx \hat{I}(d,n) = P\sum_{x\in\Phi}\tilde{\psi}(x,\mvec_x,d,n)\bigone_{\{||x|| > \delta \}}, \label{eq:intDL2}
\end{equation}
where $\tilde{\psi}(x,\mvec_x,d,n)$ is the function in the sum (\ref{eq:intDL1}) with all the $B_{x,i,j} = 1$.
%\begin{multline}
%\tilde{\psi}(x,\mvec_x,d,n) = \bigone\{W_x \leq n\} |h_{xd}|^2 l(x+S_x,d) +\\\hspace{-2mm} \sum_{i = \log n + 1}^{\log \Delta} \hspace{-2mm}\bigone\{W_x = 2^i\} \frac{ n}{2^i} \sum_{j=1}^{2^i/n} |h_{x,j,d}|^2 l(x+S_{x,j},d). \label{eq:tildephi}
%\end{multline}
The sum in (\ref{eq:intDL2}) is over an homogeneous Poisson PP of intensity $\lambda_p$ and the non-homogeneity is given by $\bigone_{\{||x|| > \delta \}}$. The LT of a Poisson PP is \cite{BB2010}:
\begin{equation}
\mathcal{L}_{\hat{I}(d,n)}(\eta) = \exp\left\{ - \lambda_p  \int\limits_{||x||>\delta}\Ex_{\mvec_x} \left[1 - e^{-\eta \tilde{\psi}(x,\mvec_x,d,n) }\right]dx \right\}. \label{eq:Laptrans}
\end{equation}
Taking the expectation over $W_x$ and over the independent unit-mean exponential fading RVs we have:
\begin{equation}
\Ex \left[e^{-\eta \tilde{\psi}(x,\mvec_x,d,n) }\right] =  \Ex\left[\frac{\prob (W_x \leq n)}{1 +l(x+S_x,d)\eta}\right]+
\sum_{i =\log n + 1}^{\log \Delta} \prod_{j = 1}^{2^i/n} \Ex\left[\frac{\prob(W_x = 2^i)}{1 + l(x+S_{x,j},d)\eta n/2^i}\right],\label{eq:ExLap1}
\end{equation}
where the remaining expectations, which cannot be computed in closed form, are over $\{S_x\}$, $\{S_{x,j}\}$.
%\begin{align*}
%\hspace{-5mm} \mathcal{L}_{\tilde{I}(z)}(s) %&= \exp\left\{ - \lambda_p p_t \int_{\R^2} \left(1 - \Ex \left[e^{-s |h_x|^2 l(x+u_x,z)  \bigone\{||x|| > 2R_c\}}\right]\right)  dx \right\} \\
%%&= \exp\left\{ - \lambda_p p_t \int_{\R^2} \Ex \left[ 1 - \frac{1}{1 + s l(x+S_x,z)\bigone\{||x|| > 2R_c\}}\right] dx \right\},\\
%&= \exp\left\{ \hspace{-1mm} - \lambda_p p_t \hspace{-1mm} \int_{||x||>\delta} \hspace{-3mm}\Ex_{S_x} \left[\frac{s l(x+S_x,z)}{1 + s l(x+S_x,z)}\right] \hspace{-1mm} dx \hspace{-.5mm}\right\} \hspace{-1mm},\hspace{-5mm}
%\end{align*}
However, it is reasonable to introduce the far field approximation that the interfering users are seen from the typical cluster as if located at the center of their cluster. 
This is because the favorable and unfavorable positions of the interferers will be approximately canceled, because they are uniformly distributed around the center. 
With this,
%\begin{equation}
%\lceil 2^i/n\rceil = \begin{cases}
%1 & \text{if }i \leq \log n,\\
%2^i/n & \text{if }\log n < i \leq \Delta,
%\end{cases} 
%\end{equation}
(\ref{eq:ExLap1}) is approximated as:
\begin{equation}
\Ex \left[e^{-\eta \tilde{\psi}(x,\mvec_x,d,n) }\right] \approx\sum_{i=0}^\Delta \frac{\prob(W_x = 2^i)}{\left(1 + \frac{l(x,d)\eta}{\lceil 2^i/n\rceil}\right)^{\lceil\frac{2^i}{n}\rceil}}. % \label{eq:ExLap1}
\end{equation}
Replacing this equation in (\ref{eq:Laptrans}),  we have:
\begin{equation}
\mathcal{L}_{\hat{I}(d,n)}(\eta) \approx \exp\left\{ - \lambda_p \sum_{i=0}^\Delta \prob(W=2^i)\int\limits_{||x|| >\delta}\frac{\left(1 + \frac{l(x,d)\eta}{\lceil 2^i/n\rceil}\right)^{\lceil\frac{2^i}{n}\rceil}-1}{\left(1 + \frac{l(x,d)\eta}{\lceil 2^i/n\rceil}\right)^{\lceil\frac{2^i}{n}\rceil}} dx \right\}.\label{eq:lapapprox1}
\end{equation}
The summation in (\ref{eq:lapapprox1}) has few terms and hence  is straightforward to implement numerically.

\section{Relevant plots and comments} \label{sec:plots} 
In this section we study the trade-off regions for strategy $\mathcal{F}^*$ (Theorem \ref{teo:servicep}), in order to study the performance attainable through D2D. We consider two PPs for the clusters centers: the type II Matérn PP, denoted as $\Phi_{HC}$, and the translated grid model, denoted as $\Phi_{TG}$. The inter-cluster attenuation model is given by (\ref{eq:interc}), with  $l(x,y) = \tilde{C} ||x-y||^{-\alpha}$, $\tilde{C}$ is a constant and $\alpha=4$. % is the path loss exponent. 
For the intra-cluster attenuation model given by $|g_{x,y}|^2$ we consider two scenarios: one in which it is the same as the inter-cluster model, and another one with lognormal shadow fading in which there may be LOS inside a cluster.

\begin{figure} [!t]
	\centering
		\begin{tikzpicture}
		\begin{semilogxaxis}[scale=1.2,tight background, 
		xlabel={$\eta$},height=6cm,width=8.7cm,
		ylabel={$\mathcal{L}^0_{I(d,n)}(\eta)$},
		ymin = 0, ymax = 1,xmin=1e5,xmax=1e9,minor x tick num=3,minor y tick num=1,
		yticklabel style={/pgf/number format/precision=1},	
		legend style = {font = \tiny,at={(0.005,0.005)},anchor = south west,inner sep = 2pt},
		legend entries = {
			$N_{m,0} = 8$ - $z = 0$ - M,
			$N_{m,0} = 8$ - $z = 0$ - (\ref{eq:lapapprox1}),
			%$N_{m,0} = 16$ - $z = 0$ - M,
			%$N_{m,0} = 16$ - $z = 0$ - T,
			$N_{m,0} = 8$ - $z = 35$ - M,
			$N_{m,0} = 8$ - $z = 35$ - (\ref{eq:lapapprox1}),
			%$N_{m,0} = 16$ - $z = 35$ - M,
			%$N_{m,0} = 16$ - $z = 35$ - T,
		}]
		\addplot[red,mark = o,mark size =1.5pt] table {FigLsim-z0slots8.dat};
		\addplot[blue,mark = *,mark size =1.5pt] table {FigLbou-z0slots8.dat};
		%
		%%%\addplot[brown,mark = diamond] table  {Dats/FigLsim-z0slots16.dat};
		%\addplot[cyan,mark = triangle*] table {Dats/FigLbou-z0slots16.dat};
		
		\addplot[magenta,mark = square,mark size =1.5pt] table {FigLsim-z35slots8.dat};
		\addplot[green,mark = diamond*] table {FigLbou-z35slots8.dat};
		
		%\addplot[pink,mark = diamond] table  {Dats/FigLsim-z35slots16.dat};
		%\addplot[orange,mark = triangle*] table {Dats/FigLbou-z35slots16.dat};
		\end{semilogxaxis}
		\end{tikzpicture} 
		\caption{Evaluation of (\ref{eq:lapapprox1}) as an approximation of the LT of the interference for a type II Matérn PP. M: Monte Carlo simulation (2000 runs each point) of the LT  of (\ref{eq:intDL1}) at $z$ by simulating the PP. $\delta/2 = R_c = 50$, $\alpha = 4$, $\lambda_u = 4 \lambda_r = 0.072$, $M=6$, $\varepsilon=0.5$. }
		\label{fig:LapCompare}

\end{figure}
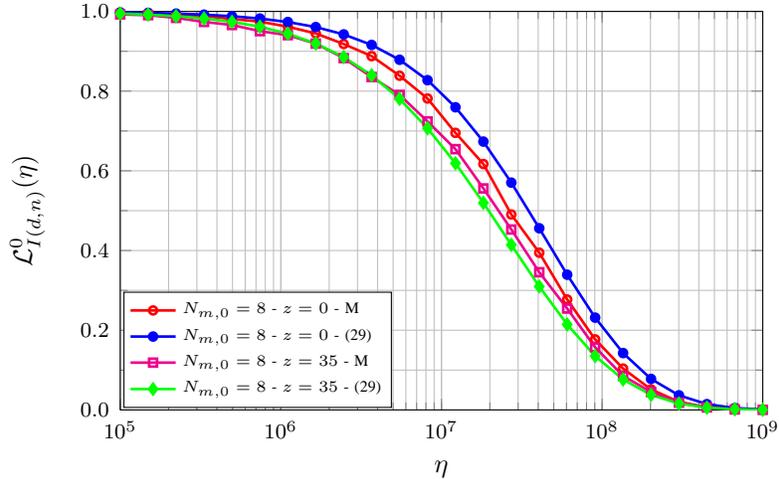

First, we consider the intra-cluster attenuation model is the same as the inter-cluster model, that is, $|g_{x,y}|^2 = |h_{x,y}|^2 l(x,y)$. We focus on the clustering PP  $\Phi_{HC}$ and the fading is Rayleigh, that is, $|h_{xy}|^2$ are unit mean exponentials. We take $P=\tilde{C}=1$, since these constants will cancel out when computing the SIR. The user densities are  $\lambda_u = 4\lambda_r = 0.012$ users/area, each caching user stores $M = 6$ videos.
In Fig. \ref{fig:LapCompare}, we plot the LT approximation (\ref{eq:lapapprox1}), when compared to the Monte Carlo simulation of the interference (\ref{eq:intDL1}) using Lemma \ref{teo:LTbou}. We evaluate this at different places in the cluster and taking a number of slots $N_{m,0} =8$, which is the most likely for the distribution of matches (the rest of the parameters are indicated in the caption).  
%We see that the approximation of the LT is very good. If we chose a value of $N_{m,0}$ which has little probability of occurring, the error will increase, because there will be a large difference between the number of slots in the cluster at the origin and the number of slots in most clusters in the network. Since this situation has a small probability of occurrence, it does not introduce a large error when we average with respect to the number of matches to compute (\ref{eq:lapapprox1}). In general, the approximation of Palm distribution of the hard-core PP by a non homogeneous Poisson PP as proposed in \cite{Haenggi_MeanInt2011} is very accurate for all network parameters. Considering the interfering users as located at the centers of the cluster introduces a larger error than the approximation on the PP.
We see that the approximation of Palm distribution of the hard core PP by a non-homogeneous Poisson PP as proposed in [17] is very accurate. Although we do not show it
here, considering the interfering users as located at the centers of the cluster introduces a larger error than the approximation on the PP.

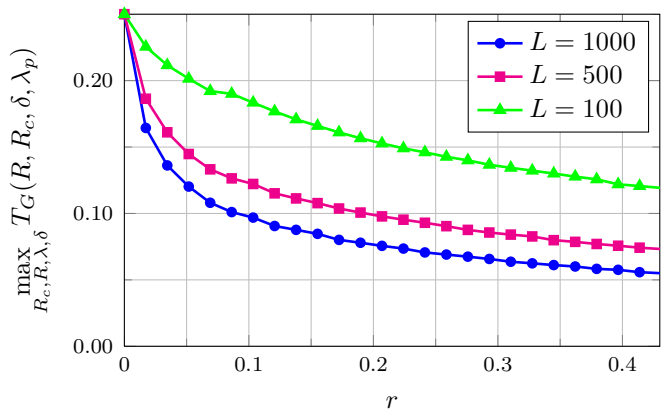
\begin{figure}[!t]
\centering
\begin{tikzpicture}[scale = 1,tight background]
\begin{axis}[height=6cm,width=8.7cm,
xlabel={$r$},
ylabel={$\displaystyle \max_{R_c,R,\lambda,\delta} T_G(R,R_c,\delta,\lambda_p)$},
ymin = 0, ymax = 0.25,xmin=0,xmax=0.43,minor x tick num=1,minor y tick num=1,
legend entries = {$L=1000$,$L=500$,$L=100$}]
\addplot[blue,mark = *,mark size =1.5pt] table {L1000-05.dat};
%\addplot[red,mark = diamond*,mark size =2pt]  table {L1000-95.dat};
\addplot[magenta,mark = square*,mark size =1.5pt] table {L500.dat};
\addplot[green,mark = triangle*] table {L100.dat};
\end{axis}
\end{tikzpicture}
\caption{
Inner bound (\ref{eq:maximiz1}) given by $(\Phi_{HC},\mathcal{F}^*)$ to the Global trade-off region (\ref{eq:TgRtoff}) for a Rayleigh fading model. 
$L$ is the library size. $\lambda_u = 0.012 = 4\lambda_r$, $M = 6$, $\gamma = 0.6$, $\alpha = 4$. $\varepsilon = 0.05$.}
\label{fig:GlobOpt}
\end{figure}

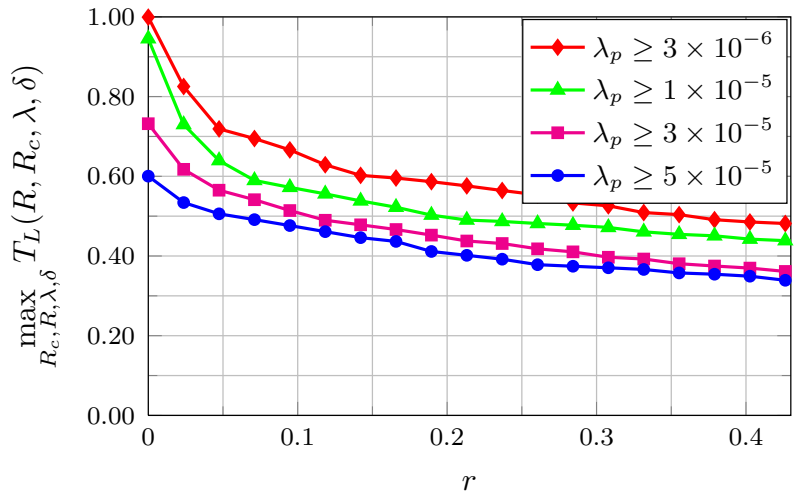
\begin{figure} [!t]
\centering
\begin{tikzpicture}[scale =1.2,tight background]
\begin{axis}[
xlabel={$r$},height=6cm,width=8.7cm,
ylabel={$\displaystyle \max_{R_c,R,\lambda,\delta}T_L(R,R_c,\lambda,\delta)$},
ymin = 0, ymax = 1,xmin=0,xmax=0.43,minor x tick num=1,minor y tick num=1,	legend style = {font = \footnotesize,at={(0.995,0.995)},anchor = north east,inner sep = 2pt},
legend entries = {
$\lambda_p \geq 3 \times 10^{-6}$,
$\lambda_p \geq 1 \times 10^{-5}$,
$\lambda_p \geq 3 \times 10^{-5}$,
$\lambda_p \geq 5 \times 10^{-5}$,}]
\addplot[red,mark = diamond*,mark size =2pt]  table {locconst-lambdap3e-006.dat};
\addplot[green,mark = triangle*] table {locconst-lambdap1e-005.dat};
\addplot[magenta,mark = square*,mark size =1.5pt] table {locconst-lambdap3e-005.dat};
\addplot[blue,mark = *,mark size =1.5pt] table {locconst-lambdap5e-005.dat};
\end{axis}
\end{tikzpicture}
	\caption{Inner bound  (\ref{eq:maximiz2}) given by $(\Phi_{HC},\mathcal{F}^*)$ to the local trade-off region (\ref{eq:TcRtoff}) for a Rayleigh fading model.
$\lambda_u = 0.012 = 4\lambda_r$, $L = 500$, $M = 6$, $\gamma = 0.6$, $\alpha = 4$.  $\varepsilon = 0.05$.} 
\label{fig:LocalOpt}
\end{figure}

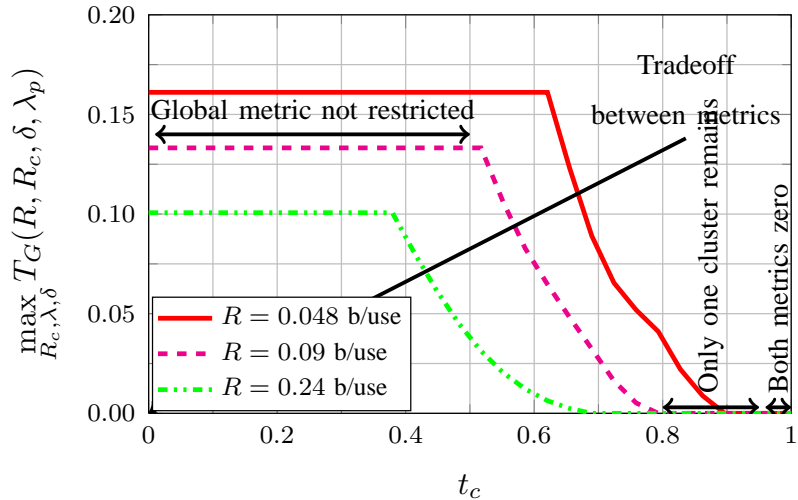
\begin{figure} [!t]
	\centering
	%\begin{center}
	\begin{tikzpicture}[scale = 1.2,tight background]
	\begin{axis}[height=6cm,width=8.7cm,
	anchor = origin,
	xlabel={$t_c$},
	ylabel={$\displaystyle \max_{R_c,\lambda,\delta} T_G(R,R_c,\delta,\lambda_p) $},
	ymin = 0, ymax = 0.2,xmin=0,xmax=1,minor x tick num=1,minor y tick num=1,
	legend style = {font = \scriptsize,inner sep = 1pt,at={(0.005,0.005)},anchor = south west,},
	legend entries = {
		$R = 0.048 $ b/use,$R = 0.09 $ b/use,$R = 0.24$ b/use,}, ]
	\addplot[red,line width = 1.5]  table {locglob-Rate0.048.dat};
	\addplot[magenta,dashed,line width = 1.5] table {locglob-Rate0.09.dat} coordinate [pos=0.55] (A) ;
	\addplot[green,dashdotdotted,line width = 1.5] table {locglob-Rate0.24.dat} ;
	%\draw[black, very thick,<-] (A)  -- (axis cs:0.71,0.12) node[anchor = south  west]{\footnotesize{\footnotesize{bet. metrics}}};
	\draw [black, very thick,<-] (A) -- (axis cs:0.836,0.138) node[above,thick,align=center,font = \footnotesize] {Tradeoff\\between metrics};
	\draw[black, very thick,<->] (axis cs:0.01,0.14) --node[above] {\footnotesize Global metric not restricted} (axis cs:0.5,0.14);
	\draw[black, very thick,<->] (axis cs:0.96,0.003) --node[rotate = 90,right] {\footnotesize{Both metrics zero}} (axis cs:1,0.003) ;
	\draw[black, very thick,<->] (axis cs:0.8,0.003) --node[rotate = 90,right] {\footnotesize{Only one cluster remains }} (axis cs:0.95,0.003) ;
	\end{axis}
	\end{tikzpicture}
	\caption{Inner bound (\ref{eq:maximiz3}) given by $(\Phi_{HC},\mathcal{F}^*)$ to the Local-Global trade-off region (\ref{eq:TgTctoff}) for a Rayleigh fading model. $R$ is the attempted rate. 
		$\lambda_r = 0.003 = \lambda_u/4$, $L = 500$, $M = 6$, $\gamma = 0.6$, $\alpha = 4$.  $\varepsilon = 0.05$.}
	\label{fig:GlobLocBal}
\end{figure}

In Figs. \ref{fig:GlobOpt}, \ref{fig:LocalOpt}, and \ref{fig:GlobLocBal} we study the inner bounds to the trade-off regions under the same setup as the previous figure. We use (\ref{eq:lapapprox1}) to approximate the LT of the interference and Monte  Carlo simulations to average over $S$ and $D$ and to estimate the distribution of $N_m$.
In Fig. \ref{fig:GlobOpt} we plot the inner bound of the global metric trade-off region (\ref{eq:TgRtoff}) achievable by $(\Phi_{HC},\mathcal{F}^*)$, changing the library size $L$, that is:
\begin{equation}
\max_{R_c,R,\lambda,\delta}  T_G(R,R_c,\delta,\lambda_p)  \text{ subject to }  \bar{R}(R,R_c,\lambda,\delta) \geq r.\! \label{eq:maximiz1}
\end{equation}
%{\color{red} Having a good approximation of the LT of the interference allows the optimization over all parameters of the model.
Optimizing over all parameters $R_c$, $R$, $\lambda$, $\delta$ implies this is the best global fraction of served requests achievable by $(\Phi_{HC},\mathcal{F}^*)$. Locally, increasing $L$ requires, on average, larger clusters  to find matches, increasing the chance of failed transmissions through path loss. This can be mitigated with a bigger cluster separation $\delta \geq 2R_c$, which reduces the density of clusters, and hence the average level of interference, but may negatively impact the density of served requests, because the clusters cover a smaller fraction of the network. A pair $(R_c, \delta)$ should be chosen to balance these effects; although not plotted, simulations show that in this setup, the optimal value was always $\delta = 2R_c$. In addition, by using Lemma \ref{teo:toffbou} and the density of the hard-core PP (Section \ref{sec:userdesc}), we can find an upper bound for the global metric of this protocol:
\begin{equation}
 T_G(R,R_c,\delta) \leq (1-e^{-\lambda \delta^2}) p_{\M} \frac{R_c^2}{\delta} \leq \frac{1}{4},
\end{equation}
for any $R$ and hence any average rate constraint. In the last inequality we used that $\delta \geq 2R_c$. 
The plot shows that when the rate is unconstrained ($r=0$ in  (\ref{eq:maximiz1})), then with $\delta=2R_c$, large $\lambda$, and $R \rightarrow 0$, the bound is achieved.
%Also, when the rate is unconstrained (take $r = 0$ in (\ref{eq:maximiz1})) then by taking $\delta = 2R_c$ and a large value of $\lambda$ (which saturates the cluster density for a given $R_c$) and a rate $R \rightarrow 0$, the bound is achieved, as the plot reflects. 
If the global metric indicates the fraction of users not requiring a downlink transmission, the plot shows that, even through this simple strategy, D2D could serve a reasonable number of requests.
The global trade-off does not guarantee a certain percentage of service within a cluster, which may be important in some scenarios: there may be many clusters with a small percentage of served requests (small $T_L$) which results in large overall benefits (large $T_G$). In Fig. \ref{fig:LocalOpt} we plot the inner bound of the local metric trade-off region  (\ref{eq:TcRtoff}) achievable by $(\Phi_{HC},\mathcal{F}^*)$, which maximizes the mean ratio of served requests:
\begin{equation}
\max_{R_c,R,\lambda_p,\delta} T_L(R,R_c,\delta,\lambda_p) \text{ subject to } \bar{R}(R,R_c,\lambda,\delta) \geq r \text{ and }\lambda_p(\delta, \lambda) \geq \lambda_t. \label{eq:maximiz2} 	
\end{equation}
%where we omitted the dependence on $(\Phi_{HC},\mathcal{F}^*)$.
The fraction of per cluster served requests, which is not bounded by the coverage percentage of cluster PP, can be much larger according to setup parameters and the value of the restrictions.
%We see that the fraction of local requests that can be served through D2D are not bounded a priori by the clustering algorithm as in the global metric, and hence a larger percentage of the demands can be served per cluster according to setup parameters and the value of the restrictions. 
%As in the global trade-off, we can use Lemma \ref{teo:toffbou} to find the value of $T_L$ when the rate is unconstrained ($r=0$), or as $r \rightarrow 0$. We use that:   $ \lambda_p \leq (4\pi R_c^2)^{-1}$,
%\begin{equation}
%\lambda_p = \frac{(1-e^{-\lambda \pi \delta^2 })}{\pi \delta^2} \leq \frac{1}{4\pi R_c^2}
%\end{equation}
%with equality obtained  as $\lambda \rightarrow \infty$ and with $2R_c = \delta$. On the other hand, as $r \rightarrow 0$ we have that $T_L(\Phi_p,\mathcal{F},R) \rightarrow \prob(\mathcal{M})$. Since $\prob(\mathcal{M})$ increases with $R_c$ and $\lambda_p$ decreases, the maximization of $T_L$ as $r\rightarrow 0$ is achieved for the maximum value of $R_c = \delta/2$ which verifies the constraint on $\lambda_p$.

In Fig. \ref{fig:GlobLocBal} we plot  the local-global metrics trade-off region  (\ref{eq:TgTctoff}) inner bound given by $(\Phi_{HC},\mathcal{F}^*)$: %, which is:
\begin{equation}
\!\max_{R_c,\lambda,\delta} T_G(R,R_c,\delta,\lambda_p) \text{ subject to } T_L(R,R_c,\delta,\lambda_p) \geq  t_c. \!\!\! \label{eq:maximiz3}
\end{equation}
We see that if  $t_c$ is small enough, then the global metric can be maximized without restriction, and in this regime, the local metric can be set as needed. A smaller local metric implies  there is a large number of clusters with a low fraction of served requests, while a larger local metric implies fewer, larger clusters with a higher level of service. If $t_c$ is larger, both constraints cannot be satisfied at the same time, and there is a trade-off between the metrics, which depends on the setup. If $t_c$ is even larger, then the maximal density of served requests becomes negligible, which implies that only the cluster at the origin remains. Finally, if the local constraint is too large it cannot be satisfied for any parameters.

After exploring the inner bounds to the trade-off regions for Rayleigh fading, we now introduce a model with shadow fading in which LOS may be present between the users in the clusters. For the intra-cluster model, we assume an indoor office model like the A1 Winner II model \cite{WinnerII}, in which the power attenuation coefficient for a transmission from $x$ to $y$ (in dB) is given by:
\begin{equation}
|g_{x,y}|^2 \text{[dB] }= C_1 \log_{10}(||x-y||) + C_2 +C_3 \log_{10}(f_c[\text{GHz}]/5)+ 5 N_w(x,y) + \chi_{x,y},
\end{equation} 
where $f_c$ is the carrier frequency, $\chi_{x,y}$ is a lognormal shadow fading coefficient of zero mean, and $N_w$ is the number of walls between $x$ and $y$. The constants $C_1, C_2, C_3$, the value of $N_w$ and the standard deviation of $\chi_{x,y}$ 
depend on whether there is LOS between the transmitter and the destination or not.
The event of LOS between $x$ and $y$ is determined according to the LOS probabilities of the Winner II A1 model \cite{WinnerII}, which is:
\begin{equation*}
\prob(\text{LOS}(x,y)) = %\begin{cases} 1 & \text{$||x-y|| \leq 2.5\text{m}$}\\
1 - 0.9 (1-(1.24-0.61 \log_{10}{||x-y||})^3)^{\frac{1}{3}}, %& ||x-y|| > 2.5\text{m}\end{cases}
\end{equation*}
when $||x-y||>5$ and one, otherwise.
The value of the constants under LOS and NLOS are:
\begin{alignat}{4}
 C_1 = 18.7,\ \  \ \ & C_2 = 46.8, \ \   \ \ & C_3 = 20 \ \ \ &\text{LOS}\\
 C_1 = 36.8, \ \  \ \ & C_2 = 43.8, \ \  \ \ & C_3 = 23 \ \ \  &\text{NLOS}.
\end{alignat}
The standard deviation of $\chi_{xy}$ are $\sigma_{\text{LOS}} = 3$ (for the LOS case) and $\sigma_{\text{NLOS}} = 6$. Since we do not define a deterministic wall distribution in the clusters, we introduce a simple model for $N_w$, which determines the number of walls as a function of the distance between the nodes:
\begin{equation}
N_w(x,y) = \begin{cases}
0 & \text{LOS}\\
1 + \lfloor(||x-y||/5-1)^+\rfloor & \text{NLOS}.
\end{cases}
\end{equation}
We assume that transmissions take place in the WiFi band, with $f_c = 2.45$GHz. Users  transmit with power $P=10^{G_tG_rP_{tx}/10}$, where $G_t = 12$dB is the transmit antenna gain, $G_r = 0$dB is the receive antenna gain, and $P_{tx} = 20$dBm. 
For the inter-cluster attenuation model, we use a variation of the B4 Winner model \cite[(3.23), part II]{WinnerII}. We assume that no LOS is possible, and  a transmission from $x$ to $y$ in different clusters to be attenuated (in dB) as:
\begin{equation}
|h_{x,y}|^2 l(x,y) \text{[dB]} = 40 \log_{10}(||x-y||) + 41 + 22.7 \log_{10}(f_c[\text{GHz}]/5) + 28 N_b(x,y) + \chi_{x,y},
\end{equation}
where $\chi_{x,y}$ is a lognormal RV with zero mean and standard deviation $\sigma=7$. $N_b=1,2,...$ takes into account the attenuation due to the clusters which are on the line between the source and the destination. This is a simple model considering a 14dB attenuation for each time the transmission penetrates or leaves a cluster between the source and the destination. 
We now evaluate the performance with this shadow fading model with possible LOS and the Rayleigh fading model without LOS (used in Figs \ref{fig:GlobOpt}, \ref{fig:LocalOpt}, and \ref{fig:GlobLocBal}) in different scenarios.  We always consider $\lambda_r= \lambda_u = 0.0278$, $L=300$ and change the number of cached videos $M$.

\begin{figure} [!t]
	\centering
	\begin{tikzpicture}
	\begin{axis}[scale=1.2,tight background,xlabel={$r $},ylabel={$T_L$},
	ymin = 0, ymax = 1,xmin=0,xmax=.25,minor x tick num=3,minor y tick num=1, height=6cm,width=8.7cm, yticklabel style={/pgf/number format/precision=1},	
	xticklabel style={/pgf/number format/precision=2},legend style = {font = \scriptsize,at={(0.995,0.995)},anchor = north east,inner sep = 2pt},
	legend entries = {
		{$M = 5, \varepsilon = 0.1$ - Rayl},
		{$M = 10, \varepsilon = 0.1$ - Rayl},
		{$M = 20, \varepsilon = 0.2$ - Rayl},
		{$M = 5, \varepsilon = 0$ - Logn},
		{$M = 10, \varepsilon = 0$ - Logn},
		{$M = 20, \varepsilon = 0.1$ - Logn},
	}]
	\addplot[red,mark = o,mark size =1.5pt] table {TLTGNLOSM5.dat};
	\addplot[blue,mark = *,mark size =1.5pt] table {TLTGNLOSM10.dat};
	
	\addplot[brown,mark = diamond] table  {TLTGNLOSM20.dat};
	\addplot[cyan,mark = triangle*] table {TLTGLOSM5.dat};
	
	\addplot[magenta,mark = square,mark size =1.5pt] table {TLTGLOSM10.dat};
	\addplot[green,mark = diamond*] table {TLTGLOSM20.dat};
	
	\end{axis}
	\end{tikzpicture}
	\caption{Fraction of mean served requests for an average rate constraint given by $(\Phi_{TG},\mathcal{F}^*)$ (inner bound to the local trade-off region (\ref{eq:TcRtoff}))
		for the lognormal shadowing and LOS model, and the Rayleigh fading model. $R_c=20$, $\delta= 50$, $\lambda_r= \lambda_u = 0.0278$, $L=300$, $\gamma= 0.4$. }
	\label{fig:TGLOScomp}
\end{figure}
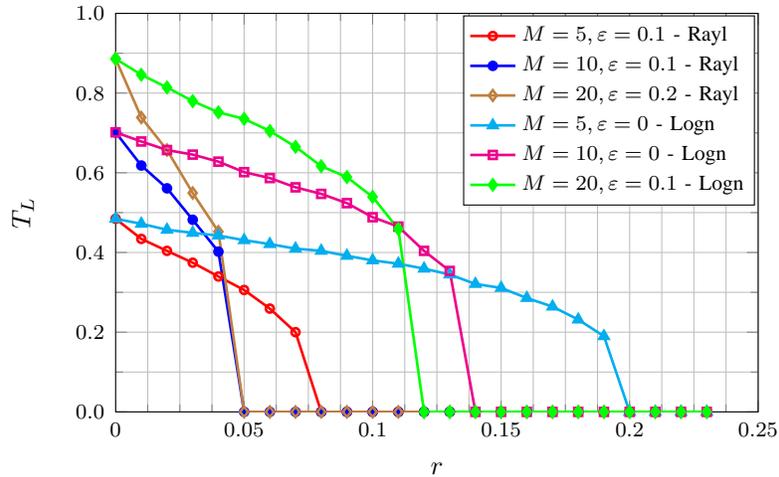

In Figs. \ref{fig:HCLOScomp} and \ref{fig:TGLOScomp} we consider both attenuation models and evaluate the local trade-off region (\ref{eq:TgRtoff}). 
For the Rayleigh model we approximate the LT of the interference using (\ref{eq:lapapprox1}), while for the shadowing model we use a Monte Carlo simulation of the PP. The value of $\varepsilon$ is chosen to maximize the local metric $T_L$ for an attempted rate $R=0.05$b/use. The number of videos cached $M$ and the fraction of dropped requests $\varepsilon$  leverage the spectral efficiency and the probability of finding videos in the cluster. A smaller value of $M$ results in a smaller fraction of served requests but allows the system to achieve a higher average rate. In Fig. \ref{fig:HCLOScomp} the cluster PP is Matérn type II; since this PP is not regular, to simplify the simulations we take $N_b=1$. We consider a cluster radius $R_c = 20$, with minimal clearance $\delta=2R_c$ (the rest of the parameters are in the caption). We can see both models provide similar results when the required average rate is small, while the model with LOS has a better performance when a higher average throughput is required. This is caused by considering the possibility of LOS and also the penetration loss to the cluster. Finally, in Fig. \ref{fig:TGLOScomp} the cluster PP is the translated grid with clearance $\delta= 50$m and cluster radius $R_c = 20$m. Each cluster could represent a block of buildings in an urban scenario. Since the process is regular, the coefficient $N_b$ between two clusters centered at $x_1$ and $x_2$ is $||x_1-x_2||_\infty/R_c$, where $||\cdot||_\infty$ is the standard infinite norm in $\R^2$. All the other parameters are the same as in Fig. \ref{fig:HCLOScomp}. In this case, the lognormal model predicts more important gains than the Rayleigh fading one. This is mainly because the Rayleigh model does not consider attenuation of the interference when penetrating the cluster, while the other one does. We see that the LOS model, which is more realistic in an urban scenario, predicts that a large number of requests could be served through D2D communications.

\begin{figure} [!t]
	\centering
	\begin{tikzpicture}
	\begin{axis}[scale=1.2,tight background,height=6cm,width=8.7cm,
	xlabel={$r $},
	ylabel={$T_L$},
	ymin = 0, ymax = 1,xmin=0,xmax=.25,minor x tick num=3,minor y tick num=1,
	yticklabel style={/pgf/number format/precision=1},	
	xticklabel style={/pgf/number format/precision=2},	
	legend style = {font = \tiny,at={(0.995,0.995)},anchor = north east,inner sep = 2pt},
	legend entries = {
		{$M = 5, \varepsilon = 0.1$ - Rayl},
		{$M = 10, \varepsilon = 0.1$ - Rayl},
		{$M = 20, \varepsilon = 0.2$ - Rayl},
		{$M = 5, \varepsilon = 0$ - Logn},
		{$M = 10, \varepsilon = 0$ - Logn},
		{$M = 20, \varepsilon = 0.1$ - Logn},
	}]
	\addplot[red,mark = o,mark size =1.5pt] table {TLHCNLOSM5.dat};
	\addplot[blue,mark = *,mark size =1.5pt] table {TLHCNLOSM10.dat};
	
	\addplot[brown,mark = diamond] table  {TLHCNLOSM20.dat};
	\addplot[cyan,mark = triangle*] table {TLHCLOSM5.dat};
	
	\addplot[magenta,mark = square,mark size =1.5pt] table {TLHCLOSM10.dat};
	\addplot[green,mark = diamond*] table {TLHCLOSM20.dat};
	
	\end{axis}
	\end{tikzpicture}
	\caption{Fraction of mean served requests for an average rate constraint  given by $(\Phi_{HC},\mathcal{F}^*)$ (inner bound to the local trade-off region (\ref{eq:TcRtoff}))
		for the lognormal shadowing and LOS model, and the Rayleigh fading model. $R_c=20$, $\delta= 2R_c$, $\lambda= 2\times10^{-4}$, $\lambda_r= \lambda_u = 0.0278$, $L=300$. $\gamma= 0.4$.}
	\label{fig:HCLOScomp}
\end{figure} 
\section{Summary and Discussion} \label{sec:discus}

We described a framework which can be used to evaluate the theoretical performance attainable through D2D. In this framework we considered a simple communication strategy, which gives an inner bound to the performance trade-off regions introduced. 
We have studied the performance of this strategy under several attenuation models.
The analysis shows that, even through our simple communication protocol, a substantial number of requests could be served through a distributed caching policy without a dedicated caching infrastructure, reducing the downlink traffic on the cellular band and a reduced load on the backhaul. Through the analysis in which we have considered a whole communications block, we have observed that the design of D2D architectures should balance the density of videos available through local caching with a proper use of time/frequency resources, so as to maximize the number of served requests. Increasing the number of cached videos will generally improve performance but choosing a transmitter randomly may imply that the average source-transmitter distance is not decreased. Therefore, we believe that the results presented in the simulation sections could be improved by considering a strategy which pairs users according to their distance. In this way, the attenuation incurred by randomly choosing a far away transmitter or by a poor choice of the cluster size could be mitigated without tuning any parameters. However, this would also required further information and coordination. Other medium access strategies which require less coordination, such as random time/frequency hopping, could also be studied to better understand how performance could be degraded if less coordination was required.

\section{Proofs}
\subsection{Proof of Lemma \ref{teo:Avreq}} \label{proof:Avreq}
For shortness we write $N_s(K)$, and using the Campbell-Mecke formula for an independently marked point PP \cite{BB2010,stochastic_geometry2009} we have:
\begin{equation*}
\Ex \left[N_s(K)\right]=\lambda_p \int_{\R^2} \hspace{-1mm} \Ex^{x} \left[ \sum_{i=1}^{N_{x,r}}\bigone_{\{x+D_{x,i} \in K\}} \bigone_{\mathcal{S}_{x,i}}\right] dx,
\end{equation*}
where $\Ex^x$ is the  Palm distribution of the PP with a cluster centered at $x$.  With all this we can write:
\begin{equation*}
\Ex \left[N_s(K)\right]=  
\lambda_p \int_{\R^2} \Ex^{0}\left[ \sum_{i=1}^{N_{0,r}}\bigone_{\{D_{0,i} \in K_{-x}\}} \bigone_{\mathcal{S}_{0,i} } \right] dx,
\end{equation*}
where $K_{-x}$ is the set obtained by shift every point in $K$ by $(-x)$. Moving the integral inside we conclude by noting that:
$\int_{\R^2} \bigone_{\{D_{0,i} \in K_{-x}\}} dx = |K|.$

\subsection{Proof of Lemma \ref{teo:toffbou}} \label{proof:toffbou}
Suppose a protocol  $(\Phi_p, \mathcal{F})$ is chosen. We can write $\Sset_{0,i} \equiv \Sset_{0,i}(R)$ to indicate the dependence of the service event with the required rate $R$. For a given realization of $\tilde{\Phi}$, decreasing $R$ cannot decrease the number of served requests, that is:
\begin{equation}
\sum_{i=0}^{N_{r,0}} \bigone_{\Sset_{0,i}(R_1) }\leq \sum_{i=0}^{N_{r,0}} \bigone_{S_{0,i}(R_2)},
\end{equation}
when $R_1 > R_2$. Taking expectation $\Ex^{0}[\cdot]$ on both sides, we can upper bound the right by taking the limit as $R_2 \rightarrow 0$. But when $R=0$ we have that: $\Sset_{0,i} = \M_{0,i}$, that is, when the required rate is $0$, a request is served whenever the video is available in the cluster, and hence, applying the monotone convergence theorem \cite{Ash2000} (with equality as $R\rightarrow 0$):
\begin{equation*}
\Ex^{0}\left[\sum_{i=0}^{N_{0,r}}  \bigone_{S_{0,i}(R)}\right] \!\leq\! \Ex^{0}\left[\sum_{i=0}^{N_{0,r}}  \bigone_{\M_{0,i}}\right]
\!=\! \Ex^0[N_{0,r}] p_{\M}.
\end{equation*}

\subsection{Proof of Lemma \ref{teo:bourate}} \label{proof:bourate}
It is straightforward to show that, given two constants $A,B >0$, the function $\phi(u) = \C \left(A/(B+u)\right)$ is convex for $u \geq 0$. Now, we rewrite the rate (\ref{eq:rach2}) in a way suitable to use the convexity of $\phi(u)$.  
%Let us denote by $I_b$ the interference from the clusters with less than or equal to $n_1$ slots as:
%$$I_b := \sum_{j=1}^{\log  n_1} I_j.$$ 
%The interference from clusters with more slots can be understood by looking at Fig. \ref{fig:syncint2}. 
%We know that a cluster with $2^kn_1$ slots will generate $2^k$ interference powers during a slot of the cluster with $n_1$ slots. Therefore for any $1 \leq k \leq \hat{k}$, $\hat{k} := \log (\Delta/n_1)$, we can index these values as:
%$$I_k(u_1,\ldots,u_k), \ \ \ \ (u_1,\ldots,u_k) \in \{1,2\}^k,$$
%We index the interferences according to number of clusters in each slot thus:
%\begin{align*}
%I_1(u_1), \ \ &u_1 = 1,2, \ \text{clusters with $2n_1$ slots}\\
%I_2(u_1,u_2), \ \ &u_1,u_2 = 1,2, \ \text{clusters with $4n_1$ slots}\\
%I_k(u_1,\ldots,u_k), \ \ &u_1,\ldots,u_k = 1,2, \ \text{clusters with $2^kn_1$ slots.}
%\end{align*}
Using (\ref{eq:Ib}) and (\ref{eq:Iav}), (\ref{eq:achrate}) is written as:
\begin{equation*}
\frac{1}{n_1} \frac{1}{2} \sum_{u_1=1}^2 \!\ldots \frac{1}{2}\! \sum_{u_{\hat{k}}=1}^2
\!\C\left(\!\frac{|g_{xy}|^2}{I_b + I_1(u_1) + \ldots + I_{u_{\hat{k}}}(u_1,\ldots u_{\hat{k}})}\right)\!,
\end{equation*}
where for shortness we  defined $\hat{k} = \log(\Delta/n_1)$.
This expression consists of  $\hat{k}$ nested convex combinations with two terms each, giving the $\Delta/n_1$ terms seen in (\ref{eq:achrate}). Notice that the $k$-th term, $I_k(u_1,\ldots,u_k)$, depends only on the summations over $(u_1,\ldots,u_k)$ and is constant for indexes $(u_{k+1},\ldots,u_{\hat{k}})$. %, which is because interference from cluster with less slots will change less often. 
We can therefore recursively use the convexity of 
$\phi(u)$ to transfer the summations inside $\C(\cdot)$, innermost to outermost. Defining $B(k,u_1, \ldots, u_k) = I_b + \sum_{i=1}^k I_i(u_1,\ldots,u_i)$. In the first step, for any set of indexes $u_1,\ldots,u_{\hat{k}-1}$ we can write:
\begin{align*}
&\hspace{-1mm}\frac{1}{2}\! \sum_{u_{\hat{k}}=1}^2
\C\!\left(\!\frac{|g_{xy}|^2}{I_b + I_1(u_1) + \ldots + I_{u_{\hat{k}}}(u_1,\ldots u_{\hat{k}})}\right)\!\\&=\frac{1}{2}\! \sum_{u_{\hat{k}}=1}^2
\C\!\left(\!\frac{|g_{xy}|^2}{B(\hat{k}-1,u_1, \ldots, u_{\hat{k}-1}) + I_{u_{\hat{k}}}(u_1,\ldots u_{\hat{k}})}\right)\!
\\
&\geq\C\!\left(\!\frac{|g_{xy}|^2}{B(\hat{k}-1,u_1, \ldots, u_{\hat{k}-1})  + \frac{1}{2}\! \sum_{u_{\hat{k}}=1}^2I_{u_{\hat{k}}}(u_1,\ldots u_{\hat{k}})}\right).
\end{align*}
Continuing this procedure $\log (\Delta/n_1)$ times, we get the  result.

\subsection{Proof of Theorem \ref{teo:servicep} }\label{proof:servicep}
We focus on the cluster at the origin, removing all subindexes $0$ (that is, $N_{0,r} \equiv N_r$, $\T_{0,i} \equiv \T_i$, etc.). For the local metric, we need to find:
\begin{equation}
\Ex^0 \left[\sum_{i=1}^{N_r} \bigone_{\Mi} \bigone_{\T_i}\right]= \Ex^0\left[ \sum_{i=1}^{N_r} \Ex^0\left[\bigone_{\Mi} \bigone_{\T_i}| N_r, N_m\right]\right].\hspace{-2mm} \label{eq:T1}
\end{equation}
In what follows, unless needed for clarity, we do not write the specific values taken by the RVs in the expectations and probabilities; for example,  $\prob^0 \left( \Mi |N_r=n_r, N_m=n_m\right) \equiv\prob^0 \left( \Mi |N_r, N_m\right)$.  
%In the last step and for the rest of the proof, the innermost expectation is over the RVs in $\mvec$ of the typical cluster that do not appear in the outer expectations. 
We have:
\begin{equation}
\Ex^0\left[\bigone_{\Mi} \bigone_{\T_i}|N_r,N_m\right] = \prob^0 \left( \Ti |N_r, N_m, \Mi\right) \prob^0 \left( \Mi | N_r, N_m \right). \label{eq:probeq2}
\end{equation}
To find $\prob^0\left( \Mi | N_r, N_m \right)$ we condition on $(N_u,\tilde{\avec})$, so:
\begin{equation}
\prob^0 \left( V_i \in \tilde{\avec}|N_r, N_m,N_u,\tilde{\avec}\right) = \frac{\binom{N_r-1}{N_m-1}}{\binom{N_r}{N_m}} = \frac{N_m}{N_r}. \label{eq:Ber1}
\end{equation}
To prove (\ref{eq:Ber1}) we consider that once $\tilde{\avec}$ is fixed, the event $V_i \in \tilde{\avec}$ can be interpreted as a success in a Bernoulli trial. Then, we are asking for the probability of a success on the $i$-th trial out of $N_r$ Bernoulli trials given that there were a total of $N_m$ successes. %The binomial coefficient in the denominator is the number of ways to assign the $N_m$ successes in the $N_r$ trials, and the numerator is the number of ways to assign $N_m-1$ successes among $N_r-1$ trials, given that the $i$-th trial is a success. 
Using (\ref{eq:Ber1}), we have:
\begin{equation}
\prob^0 \left(\M_{i}|N_r,N_m\right) = \Ex^0 \left[\frac{N_m}{N_r} \bigg|N_r,N_m\right] \nonumber= \frac{N_m}{N_r}. \label{eq:pNmNr}
\end{equation}
%\begin{equation}
%\prob^0 \left(\M_{i}|N_m,N_r\right) = \Ex^0_{N_u} \left[\Ex^0_{\tilde{\avec}} \left[ \frac{N_m}{N_r} |N_r,N_m,N_u\right]|N_r,N_m \right] \nonumber= \frac{N_m}{N_r}. \label{eq:pNmNr}
%\end{equation}
We now only need to find $\prob^0 \left( \Ti |N_r, N_m, \Mi\right)$ in  (\ref{eq:probeq2}).
The event $\Ti$ % in (\ref{eq:probeq2}) 
can be written as $\T_i = \Oi(W(N_m)) \cap \mathcal{P}_i,$
where $\mathcal{P}_i$ indicates the user was scheduled for a transmission and $\Oi(W(N_m)) = \{\text{Tx successful for user }i\}$.  In what follows we do not write the dependence of $W$, $W_L$ and $W_H$ with $N_m$ (that is, we write $W(N_m) \equiv W$, etc.). When $N_m$ is such that $W(N_m) = W_H(N_m)$, the user is always scheduled, while when $W(N_m) = W_L(N_m)$ the user may not be scheduled. So we may write:
\begin{multline}
\prob\left(\T_i|N_r, N_m, \Mi\right)=
\prob^0\left(\Oi(W_H)|N_r, N_m,\Mi\right) \bigone_{\{W= W_H\}}+
\prob^0\left(\Oi(W_L)| \mathcal{P}_i,N_r, N_m, \Mi\right)
\\\times\prob^0\left(\mathcal{P}_i|N_r, N_m, \Mi\right) \bigone_{\{W = W_L\}}. \label{eq:POi1}
\end{multline}
%\begin{align}
%\prob^0\left(\Oi(W)\cap \mathcal{P}_i|N_r, N_m, \Mi\right) &= 
%\end{align}
%Thus, for $\prob^0 \left( \Ti |N_r, N_m, \Mi\right)$ in (\ref{eq:probeq2}), we can write:
%\begin{multline}
%\prob^0\left(\Ti|N_r, N_m, \Mi\right) % = \prob\left(\Oi(W), \mathcal{P}_i|N_r, N_m, \Mi\right) \\
%= \prob^0\left(\Oi(W), \mathcal{P}_i,W = W_H|N_r, N_m, \Mi\right) \\+ \prob^0\left(\Oi(T), \mathcal{P}_i,W = W_L|N_r, N_m, \Mi\right). \label{eq:cuenta2}
%\end{multline}
%Since $\{W(N_m) = W_H(N_m)\} \subset \mathcal{P}_i$, when $N_m$ 
%\begin{align}
%\prob^0\left(\Oi(W), \mathcal{P}_i,W = W_H|N_r, N_m, \Mi\right) &=
%\prob^0\left(\Oi(W), W = W_H|N_r, N_m, \Mi\right) \\
%&= \prob^0\left(\Oi(W_H)|N_r, N_m,\Mi\right)\bigone_{\{W = W_H\}}. \label{eq:POi1}
%\end{align}
%When $N_m$ is such that $W(N_m) = W_L(N_m)$ the user may not be scheduled:
%\begin{multline}
%\prob^0\left(\Oi(W), \mathcal{P}_i|N_r, N_m, \Mi\right) = 
%\prob^0\left(\Oi(W_L)| \mathcal{P}_i,N_r, N_m, \Mi\right) \times \\
%\!\!\prob^0\left(\mathcal{P}_i|N_r, N_m, \Mi\right) \bigone_{\{W = W_L\}}. \label{eq:POi2}
%\end{multline}
%\vspace{-5mm}
By a similar argument to the one used in (\ref{eq:Ber1}), when $N_m$ is such that $W=W_L$ we have:
\begin{equation}
\prob^0\left(\mathcal{P}_i|N_r, N_m, \Mi\right) = \frac{W_L(N_m)}{N_m}. \label{eq:cuenta}
\end{equation}
Now we need to find the probability of $\Oi(W)$ appearing in both terms in (\ref{eq:POi1}). For shortness, let us define the vector of RVs $Z\triangleq(N_u,N_r,N_m,\tilde{\avec})$. We have:
\begin{equation}
\prob^0\left(\Oi(W_L)|\mathcal{P}_i,N_r, N_m,\Mi\right)=
\Ex^0\left[ \sum_{v \in \tilde{\avec}} \prob^0(\Oi(W_L), V_i=v|\mathcal{P}_i,\Mi,Z)|N_r, N_m\right]. \label{eq:probeq0}
 \end{equation}
%For the first equality it is straightforward to verify  that $(N_u,\tilde{\avec})$ conditioned on $(N_r,N_m)$ is independent of $\M_i$. 
%	Also in the first equality, it is important to mention that, since we are conditioning on $\M_i$, we only need to add over the $V_i \in \tilde{\avec}$. 
Notice that, we only need to add over the $v \in \tilde{\avec}$ because we are conditioning on $\M_i$. This guarantees that in the following step we will not condition with respect to an event of zero probability. Now, since there is a match and only one transmitter will serve the request we have:
\begin{align}
\prob^0(\Oi(W_L)|V_i,\mathcal{P}_i,  \Mi,Z)
%&= \hspace{-2mm} \sum_{c \in \Aset(\tilde{\avec},V_i)} \hspace{-2mm} \prob^0(\Oi(W_L),C_i = c| V_i,\mathcal{P}_i, N_r, N_m, N_u, \Mi,\tilde{\avec}) \nonumber\\
&=\sum_{c \in \Aset(\tilde{\avec},V_i)} \hspace{-2mm}  \frac{\prob^0(\Oi(W_L)| C_i = c,V_i,\mathcal{P}_i, \Mi,Z) }{\#\Aset(\tilde{\avec},V_i)}  \\
&= \prob^0(R_a(W_L,S,D_i)> R| N_r, N_m, N_u), \label{eq:probeq1}
\end{align}
where $S$ and $D$  follow the distribution of any user in the cluster (uniform) and $R_a$ is given by (\ref{eq:rach2}). In the last step we use that the event of  failed transmission to a user depends only on the number of slots in the block (that is, on $N_m$) and on the number of requests and caching users (otherwise the transmission may not be well defined).
%In the last step we applied (\ref{eq:Ffact1}) and (\ref{eq:distC}) so:
%\begin{gather}
%\prob^0(C_i = c| V_i, N_r, N_m, N_u, \Mi,\tilde{\avec}) = \frac{1}{\#\Aset(\tilde{\avec},V_i)}\\
%\prob^0(\Oi(W_L)| V_i, N_r, N_m, N_u, \Mi,\tilde{\avec},C_i = c) = 
%\prob^0(R_a(W_L,S_c,D_i) > R |N_r,N_m,C_i=c).
%\end{gather}
In this last step, the distribution of $S$ and $D_i$ are the same for any $c$, $i$. Thus, noticing that (\ref{eq:probeq1}) does not depend on $V_i$ and replacing it in  
(\ref{eq:probeq0}) we can sum over $V_i$. This sum yields one, because we are conditioning on a match and on $\tilde{\avec}$. Thus we obtain: 
%\begin{equation}
%\prob^0 \left( \Oi(W_L) |N_r, N_m, \Mi, \mathcal{P}_i\right) =  \prob^0(R_a(W_L,S,D_i) > R|N_m,N_r). \label{eq:probeq3}
%\end{equation}
%Thus, going back to (\ref{eq:T1}) and plugging (\ref{eq:pNmNr}) and (\ref{eq:probeq3}) we find:
%\begin{multline}
%T(\Phi_p,\mathcal{F}^*,R)= \\
%\Ex_0^! \left[\Ex_{N_m} \left[N_m \prob(\C(\textrm{SIR}(S,D)) > N_m R|N_m) \right]\right]. \label{eq:T2}
%\end{multline}
%This probability is of $S$, $D$ and the fading coefficient $|g_{SD}|^2$, which are independent of $N_m$ so: % we have:
\begin{align}
\prob^0 \left( \Oi(W_L) |N_r, N_m, \Mi, \mathcal{P}_i\right) &=  \prob^0(R_a(W_L,S,D_i) > R|N_m,N_r)
% &=\Ex_{S,D_i}^0 \left[\prob^0\left(|g_{S,D_i}|^2 > I(D_i,N_m) W_L(N_m) R\bigg|N_m,N_r,S,D_i\right)|N_m,N_r\right] \nonumber\\
\\&=\Ex^0 \left[\bar{F}_{|g_{S,D_i}|^2|S,D_i}\left(I(D,N_m) W_L(N_m) R \right)|N_m,N_r\right]. \label{eq:T3}
\end{align}
%$\bar{F}_{|h_{S,D}|^2}$ is the complementary distribution function of $|h_{S,D}|^2$ and 
$I(D,N_m)$ is the interference at $D$ (\ref{eq:intDL1}). With this we can find $\prob^0\left(\Oi(W_L)| \mathcal{P}_i,N_r, N_m, \Mi\right)$ in (\ref{eq:POi1}).
Following the same procedure we find $\prob^0\left(\Oi(W_H)|N_r, N_m,\Mi\right)$ in (\ref{eq:POi1}), which concludes the proof.	
% The same can be done using $W_L(N_m)$, which together with (\ref{eq:cuenta}) gives (\ref{eq:POi2}). Finally, replacing (\ref{eq:POi1}) and (\ref{eq:POi2}) in (\ref{eq:cuenta2}), then (\ref{eq:cuenta2}) and (\ref{eq:pNmNr}) in (\ref{eq:T1}) we get the desired result. 
%For unit mean Rayleigh fading $\bar{F}_{|h_{S,D}|^2}(u) = \exp(-u)$.
Similarly we find the average rate:
%The proof of the average rate follows on the same lines:
\begin{equation}
\bar{R}(\mathcal{F},R)=R \ \Ex^0\left[ \sum_{i=1}^{N_{r}} \frac{\bigone_{\Sset_{i}}}{ \sum_{j=1}^{N_{r}} \bigone_{\M_{j} \cap \mathcal{P}_{j}} }  \right]
=R \ \Ex^0\left[ \sum_{i=1}^{N_{r}} \frac{\bigone_{\M_{i} \cap \mathcal{P}_{i}} \Ex^0[\bigone_{\mathcal{O}_i}|\M_{i}, \mathcal{P}_{i},N_r,N_m]}{ \sum_{j=1}^{N_{r}} \bigone_{\M_{j} \cap \mathcal{P}_{j}} }  \right]. %\label{eq:defRbar}
\end{equation}
In the last step the conditional expectation does not depend on $i$ and is the same as (\ref{eq:T3}) with $W(N_m)$ instead of $W_L(N_m)$. 

\subsection{Proof of Lemma \ref{teo:LTbou}}\label{proof:LTbou}
%We review the notion of second-order stochastic dominance. %\cite{Wolfstetter1999}. 
If $X,Y$ are real RVs, $X$ second-order stochastically dominates\cite{Wolfstetter1999} $Y$, written $X\geq_{2} Y$ if, %for all $u>0$:
%$$ \int_0^u F_X(x) dx \leq \int_0^u F_Y(y) dy.$$
%This definition is equivalent to asking that 
for any monotone increasing concave function $\phi(u)$:
%\begin{equation}
$\Ex[\phi(X)] \geq \Ex[\phi(Y)].$ % \label{eq:def2ssd}
%\end{equation}
We have \cite{LiWong99}:

\begin{theorem}
If $\{X_1,\ldots X_n\}$ and $\{Y_1,\ldots,Y_n\}$ are sets of independent real RVs and $\{\beta_i\}_{i=1}^n$ are non-negative real numbers:
\begin{equation}
X_i \geq_2 Y_i \ \forall i \iff \sum_{i=1}^n \beta_i X_i \geq_2 \sum_{i=1}^n \beta_i Y_i, \label{eq:teossd1}
\end{equation}	
%Furthermore if $\sum_{i=1}^n \beta_i = 1$, then:
\begin{equation}
\frac{1}{n} \sum_{i= 1}^n X_i \geq_2 \sum_{i=1}^n \beta_i X_i, \text{ when $\sum_{i=1}^n \beta_i = 1$}.\label{eq:teossd2}
\end{equation}
\end{theorem}
%\begin{itemize}
%\item If $\{X_1,\ldots X_n\}$ and $\{Y_1,\ldots,Y_n\}$ are sets of independent real RVs and $\{\beta_i\}_{i=1}^n$ are non-negative real numbers:
%\begin{equation}
%X_i \geq_2 Y_i \ \forall i \iff \sum_{i=1}^n \beta_i X_i \geq_2 \sum_{i=1}^n \beta_i Y_i. \label{eq:teossd1}
%\end{equation}
%\item If $\{X_1,\ldots X_n\}$ and $\{\beta_i\}_{i=1}^n$ are as above and $\sum_{i=1}^n \beta_i = 1$:
%\begin{equation}
% \frac{1}{n} \sum_{i= 1}^n X_i \geq_2 \sum_{i=1}^n \beta_i X_i.\label{eq:teossd2}
%\end{equation}
%\end{itemize}
We now consider the interference (\ref{eq:intDL1}) generated only by the clusters with centers inside $\B(0,\rho)$:
\begin{equation}
I_\rho (d,n_1) = \sum_{x\in (\Phi\cap \mathcal{B}(0,\rho))\setminus\{0\} } \psi(x,d,\mvec_x,n_1), \label{eq:intred1}
\end{equation}
where $\psi(x,d,\mvec_x,n_1)$ is the function in the sum in (\ref{eq:intDL1}).
This sum has a finite number of terms almost surely. 
%Additionally, let $\tilde{\psi}(x,d,\mvec_x,n_1)$ be given by (\ref{eq:tildephi}), that is, identical to $\psi(x,d,\mvec_x,n_1)$ but with all the $\{\B_{x,i,j}\}$ equal to $1$.  
Now, we condition on the cluster centers $\Phi$, on the number of slots $\{W_x\}$ and on the variables $\{B_{x,i,j}\}$ of all the clusters. Then the only randomness in the sum (\ref{eq:intred1}) comes from the fading coefficients and the positions of the users around the cluster centers, which are all independent RVs.
With this, for a cluster with $W_x = 2^i > n_1$ and using (\ref{eq:teossd2})  we have:
\begin{multline*}
\psi(x,d,\mvec_x,n_1)= \frac{n_1}{2^i} \sum_{j=1}^{2^i/n_1} \hspace{-.5mm} B_{x,j,i} |h_{x,j,d}|^2 l(x+r_j,d) 
\geq_2\\\frac{n_1}{2^i} \sum_{j=1}^{2^i/n_1} \hspace{-.5mm} |h_{x,j,d}|^2 l(x+r_j,d)= \tilde{\psi}(x,d,\mvec_x,n_1)
\end{multline*}
where $\tilde{\psi}(x,d,\mvec_x,n_1)$ is the same $\psi(x,d,\mvec_x,n_1)$ but has $B_{x,i,j} = 1$.
Keeping the conditioning, and summing over all the clusters which have more that $n_1$ slots and using (\ref{eq:teossd1}):
\begin{equation}
\sum_{x\in (\Phi\cap \mathcal{B}(0,\rho))\setminus\{0\} } \tilde{\psi}(x,d,\mvec_x,n_1) \bigone_{\{W_x >n_1\}} \geq_2
\sum_{x\in (\Phi\cap \mathcal{B}(0,\rho))\setminus\{0\} } {\psi}(x,d,\mvec_x,n_1) \bigone_{\{W_x >n_1\}}.
\end{equation}
With the conditioning, the interference from clusters such that $W_x\leq n_1$ is independent of the ones such that $W_x >n_1$, so using (\ref{eq:teossd1}) once more get:
%\begin{equation}
$\sum_{x\in (\Phi\cap \mathcal{B}(\rho))\setminus\{0\} } \tilde{\psi}(x,d,\mvec_x,n_1) \geq_2 I_\rho (d,n_1).$
%\end{equation}
Using the definition of stochastic dominance with $\phi(u) = -e^{-s u}$, $s >0$, which is concave and increasing in $u$. Averaging over the conditioned RVs we get the desired result but for $I_\rho(d,n_1)$ instead of the full interference. We conclude by letting $\rho\rightarrow \infty$ and using monotonicity arguments.

\bibliographystyle{IEEEtran}
\bibliography{IEEEabrv,d2dcachin}

\end{document}